\documentclass{ieeeaccess}
\usepackage{cite}
\usepackage{amsmath,amssymb,amsfonts}
\usepackage{algorithmic}
\usepackage{graphicx}
\usepackage{textcomp}

\usepackage{subfigure}
\usepackage{tabularx}
\usepackage{booktabs}
\usepackage{verbatim}
\usepackage{multirow}
\usepackage[dvipsnames]{xcolor}

\definecolor{myblue}{RGB}{0, 0, 210}

\newlength{\xfigwd}
\newcolumntype{C}{>{\setlength\hsize{\hsize}\centering\arraybackslash}X}

\newcommand{\modify}[1]{\color{black}{#1}\color{black}{}}

\newcommand{\etal}{\emph{et al.}}

\def\BibTeX{{\rm B\kern-.05em{\sc i\kern-.025em b}\kern-.08em
    T\kern-.1667em\lower.7ex\hbox{E}\kern-.125emX}}

\begin{document}
\history{Date of publication xxxx 00, 0000, date of current version xxxx 00, 0000.}
\doi{10.1109/ACCESS.2021.DOI}

\title{IceNet for Interactive Contrast Enhancement}
\author{\uppercase{Keunsoo Ko}, \IEEEmembership{Student Member, IEEE}
\uppercase{, and Chang-Su Kim}, \IEEEmembership{Senior Member, IEEE}}
\address{School of Electrical Engineering, Korea University, Seoul 136-701, South Korea}
\tfootnote{This work was supported by the National Research Foundation of Korea (NRF) grant
funded by the Korea government (MSIP) (No.~NRF-2018R1A2B3003896 and No. NRF-2021R1A4A1031864).}

\markboth
{Ko and Kim: IceNet for Interactive Contrast Enhancement}
{Ko and Kim: IceNet for Interactive Contrast Enhancement}

\corresp{Corresponding author: Chang-Su Kim (e-mail: changsukim@korea.ac.kr)}

\begin{abstract}
A CNN-based interactive contrast enhancement algorithm, called IceNet, is proposed in this paper, which enables a user to adjust image contrast easily according to his or her preference. Specifically, a user provides a parameter for controlling the global brightness and two types of scribbles to darken or brighten local regions in an image. Then, given these annotations, IceNet estimates a gamma map for the pixel-wise gamma correction. Finally, through color restoration, an enhanced image is obtained. The user may provide annotations iteratively to obtain a satisfactory image. IceNet is also capable of producing a personalized enhanced image automatically, which can serve as a basis for further adjustment if so desired. Moreover, to train IceNet effectively and reliably, we propose three differentiable losses. Extensive experiments demonstrate that IceNet can provide users with satisfactorily enhanced images.
\end{abstract}


\titlepgskip=-15pt

\maketitle

\section{Introduction}

\begin{figure*}
    \centering\hfill
    \subfigure []{\includegraphics[width=0.248\linewidth]{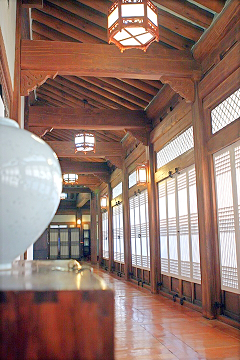}}\hfill
    \subfigure []{\includegraphics[width=0.248\linewidth]{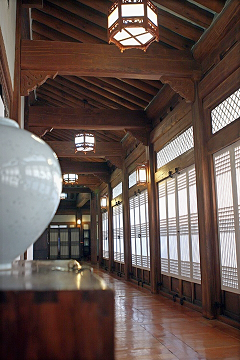}}\hfill
    \subfigure []{\includegraphics[width=0.248\linewidth]{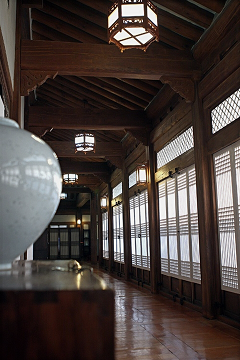}}\hfill
    \subfigure []{\includegraphics[width=0.248\linewidth]{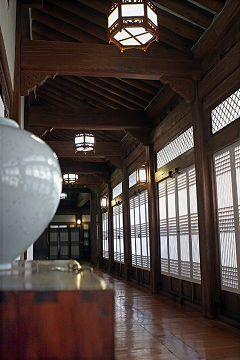}}\hfill
    \vspace{-0.1cm}
    \caption{Four different CE results of the same image using IceNet. Each result can be interpreted differently: (a) a warm hallway with a high exposure level, (b) a cool hallway with a medium exposure level, and (c) and (d) dark hallways with the same low exposure level. In (d), the foreground table and ceramic are highlighted with local brightening.}
    \label{fig:intro}
\end{figure*}

Despite recent advances in imaging technology, photographs often fail to represent scene details faithfully due to challenging factors, which include non-uniform exposure, short shutter cycle, and weak ambient light. Such photographs exhibit contrast distortions, color fading, and low intensity; especially, abnormal light conditions would distort colors, texture, and objects considerably, thereby degrading the visual experience. Contrast enhancement (CE) techniques can alleviate these problems.

Many attempts have been made to develop effective CE techniques \cite{arici2009histogram,wang2009real,celik2011contextual, lee2013contrast, guo2017lime,yang2018adaptive}. A simple but efficient approach is to use parametric curves for transformation functions from input to output pixel values\modify{\cite{wang2009real,yuan2012automatic, huang2012efficient,yang2018adaptive,srinivas2021context}}. For example, a power law is a well-known parametric curve for gamma correction~\cite{gonzalez}. These parametric curves produce promising results, but it is challenging to find reliable parameters, which are effective for diverse images. Recently, with the success of convolutional neural networks (CNNs) in the field of low-level vision \cite{goodfellow2014generative,gatys2016image,zamir2020cycleisp,ko2021light}, CNN-based CE methods also have been proposed, \modify{yielding outstanding performance \cite{liu2021benchmarking}}.

\modify{Many researches have been carried out to perform CE automatically. However, CE is a subjective process because people have different preferences for contrast. Note that} contrast plays different roles in image interpretation. For example, a viewer uses contrast to attract attention; an artist conveys emphasis through contrast; a graphic designer employs contrast to tell the eyes where to go \cite{grundland2006interactive}. Thus, there is no single definite way for enhancing contrast. Figure~\ref{fig:intro} illustrates that there can be many ways to enhance the contrast of the same image. Professional software, such as Photoshop, provides many tools to adjust contrast according to personal preference, but using such tools takes much effort. In this paper, we propose the first CNN-based interactive CE algorithm to enable a user to adjust image contrast easily and adaptively.

For interactive CE, we propose IceNet, which can enhance contrast in both under- and over-exposed regions of an image after accepting annotations from a user. Specifically, a user provides a parameter for controlling the global brightness and two types of scribbles to darken or brighten local regions in an input image. Then, we feed the input image and the annotations to IceNet, which generates a gamma map for the pixel-wise gamma correction. \modify{More specifically, we feed the scribbles, as well as the input image, to a feature extractor, yielding a feature map. Simultaneously, we produce a driving vector from the parameter for the global brightness. Then, using the feature map and the driving vector, we obtain a gamma map, which is generated adaptively according to the user annotations. Finally, we restore an enhanced image through color restoration, in which we adjust only the luminance component of the input image while preserving the chrominance components.} Also, for user's convenience, we provide an initially enhanced image and allow the user to further enhance it interactively. Moreover, to train IceNet effectively and reliably, we propose three differentiable losses. Experimental results demonstrate that IceNet not only provides users with satisfactory images but also outperforms the state-of-the-art algorithms qualitatively and quantitatively. It is strongly recommended to watch the supplemental video for a real-time demo of IceNet.

To summarize, this work has the following main contributions:
\vspace{0.2cm}
\begin{itemize}
  \itemsep2.5mm
  \item \modify{We propose the first CNN-based interactive CE algorithm, called IceNet, which is capable of yielding enhanced images either adaptively according to user preference or automatically without interaction.}
  \item \modify{We train IceNet with the proposed three differentiable loss functions, thereby enabling interaction with users and yielding pleasingly enhanced images.}
  \item \modify{With various experimental results, we demonstrate that IceNet can provide satisfactory results to users, outperforming conventional algorithms meaningfully.}
\end{itemize}

\vspace{0.2cm}
\section{Related Work}

\modify{The objectives of enhancement are closely related but different between CE \cite{kumar2020contrast,liu2021benchmarking}, color enhancement \cite{kim2020pienet}, dehazing \cite{kim2013optimized,zheng2020image,zhu2020novel}, and detail enhancement \cite{hao2016image}.} This section briefly reviews CE techniques.

\subsection{Traditional methods}
Traditional CE techniques can be classified into three categories: histogram methods, parametric curve methods, and retinex methods. First, histogram equalization~\cite{gonzalez} attempts to make the output histogram as uniform as possible. It increases contrast effectively at low computational complexity, but it may over-enhance an image, resulting in contrast over-stretching, noise amplification, and contour artifacts. Various histogram methods \cite{arici2009histogram,celik2011contextual,lee2012contrast, lee2013contrast,lim2017contrast} have been developed to overcome these problems. Second, parametric curve methods, such as gamma correction and logarithm mapping, use parametric curves as transformation functions between input and output pixel values. \modify{Many parametric curves are available for CE \cite{wang2009real,yuan2012automatic,huang2012efficient,yang2018adaptive,srinivas2021context}}. Among them, gamma correction has been extensively used not only for CE but also to match different dynamic ranges between imaging devices.  Third, assuming that an image can be decomposed into reflection and illumination, some CE algorithms \cite{wang2013naturalness,fu2016weighted,guo2017lime,li2018structure} have been developed based on retinex theory \cite{land1971lightness}. These conventional methods produce promising results in some cases. However, their performance usually depends on careful parameter tuning; it is difficult to find reliable parameters effective for diverse input images.

\begin{figure*}[t]
\centering
\includegraphics[width=0.9\linewidth]{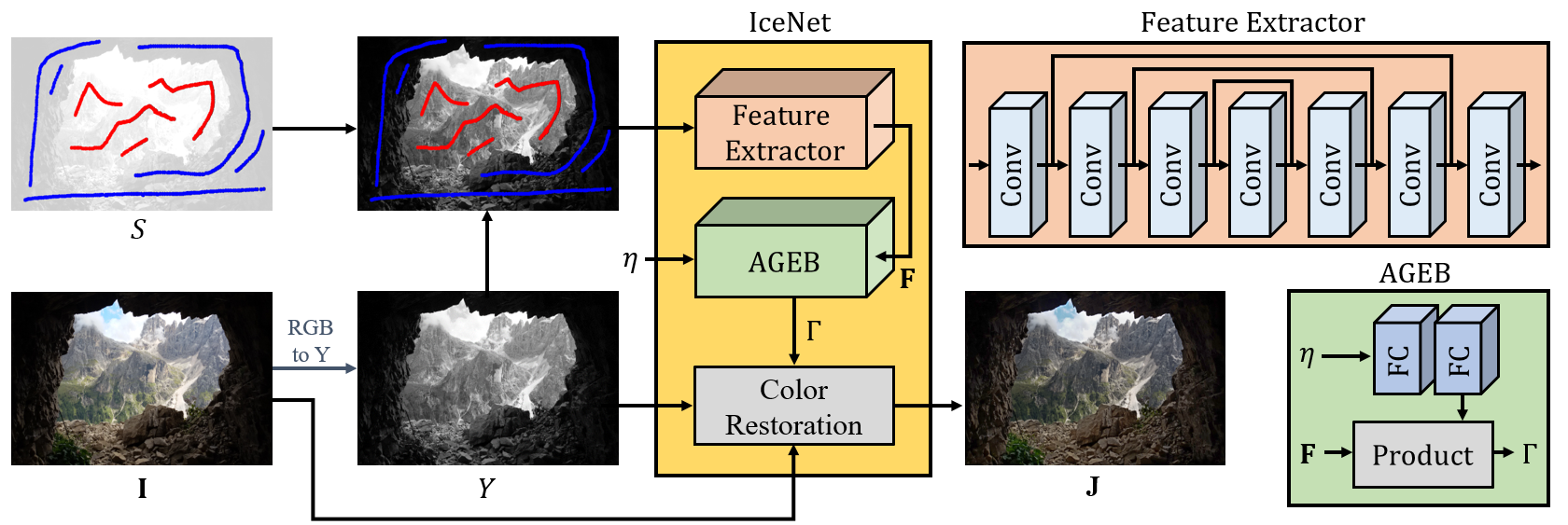}
\caption{
Illustration of the proposed algorithm. A user provides an exposure level $\eta$ and a scribble map $S$, where red or blue scribbles are for darkening or brightening local regions in an input image $\mathbf{I}$, respectively. Then, IceNet generates an enhanced image $\mathbf{J}$. The user may provide annotations iteratively, until a satisfactory image is obtained. Please see the supplemental video for a real-time demo of the interactive enhancement.}
\label{fig:overview}
\end{figure*}

\subsection{CNN-based methods}
Recently, many CNNs have been designed for CE. \modify{Most CNN-based methods \cite{lore2017llnet,li2018lightennet,park2018dual,wei2018deep, guo2019a,hu2020an,zhu2020learning,zhang2021beyond}} are trained on paired datasets, composed of pairs of low and high contrast images. Each pair is captured from the same scene, or a low contrast image is synthesized from a high contrast one. However, it is hard to capture the same scene twice because of moving objects. On the other hand, synthesized low contrast images may not be photo-realistic. Because of these difficulties, some methods \cite{kim2019low,ma2021retinexgan,jiang2021enlightengan} adopt generative adversarial networks \cite{goodfellow2014generative} and train their networks using unpaired datasets, consisting of low and high contrast images captured from different scenes. They, however, should select unpaired images carefully. To avoid this cumbersome process, Guo \etal \cite{guo2020zero} proposed a self-supervised learning scheme, which needs only low contrast images for training. However, all these CNN-based methods are not adaptive and cannot satisfy diverse user preferences.

\subsection{Interactive methods}
For interactive CE, professional software provides tools, but using these tools takes a lot of effort and training. To reduce such effort, simple interactive methods have been proposed. Stoel \etal \cite{stoel1990interactive} proposed an interactive histogram equalization scheme to enhance details of medical images. Their method allows a user to specify a region of interest (RoI) and applies the equalization to the region. Grundland and Dodgson \cite{grundland2006interactive} proposed an interactive tone adjustment method.  When a user selects key tones on an image, it preserves those tones but adjusts the other tones, while maintaining the overall tonal balance. Lischinski \etal \cite{lischinski2006interactive} and Dodgson \etal \cite{dodgson2009contrast} also proposed interactive methods, in which a user specifies RoIs for local CE. These interactive methods~\cite{stoel1990interactive, grundland2006interactive, lischinski2006interactive, dodgson2009contrast} are based on traditional image processing. In contrast, to the best of our knowledge, the proposed IceNet is the first CNN-based interactive CE algorithm.

\section{Proposed Algorithm}

To enable a user to adjust image contrast easily, we develop IceNet that accepts simple annotations and yields an enhanced image according to his or her preference.

\subsection{Interactive Contrast Enhancement}
\label{ssec:ice}

Figure~\ref{fig:overview} illustrates the proposed algorithm. By inspecting an input image $\mathbf{I}$, a user provides an exposure level $\eta \in [0, 1]$ for controlling the global brightness and two types of scribbles. Red and blue scribbles, respectively, mean that the user wants to darken and brighten the corresponding local regions. They are recorded as $-1$ and $1$ in the scribble map $S$, respectively, while the other pixels are assigned $0$. We convert the RGB color image $\mathbf{I}$ into the YCbCr space and adjust the luminance component $Y$ only, while preserving the chrominance components \cite{lee2013contrast}. Then, we estimate a gamma map $\Gamma$ for the pixel-wise gamma correction of $Y$. Finally, through the color restoration, we obtain an enhanced image $\mathbf{J}$.

\subsubsection{Gamma estimation}
Gamma correction is widely used for CE~\cite{gonzalez}. It is important to select an appropriate gamma value: a small gamma less than 1 brightens an under-exposed region, whereas a large gamma bigger than 1 darkens an over-exposed region. The gamma value also should be selected by considering personal preference.

We determine a gamma value for each pixel in an image, based on user preference, through the feature extractor and the adaptive gamma estimation block (AGEB) in Figure~\ref{fig:overview}. The feature extractor consists of seven convolutional layers with concatenated skip connections, which improve the information flow between layers. We concatenate $Y$ and $S$ to form the input to the feature extractor, which yields a feature map $\mathbf{F}$. Then, given the exposure level $\eta$ and the feature map $\mathbf{F}$, AGEB generates a gamma map $\Gamma$. More specifically, from $\eta$, AGEB produces a driving vector $\mathbf{w}$ via two fully-connected layers. Then, for each pixel $\mathbf{x}$, it predicts the pixel-wise gamma value $\Gamma(\mathbf{x})$ by
\begin{equation}
\label{eq:gamma_map}
\Gamma(\mathbf{x}) = 10\varphi(\langle\mathbf{F(x)}, \mathbf{w}\rangle)
\end{equation}
where $\langle\cdot{,}\cdot\rangle$ is the inner product and $\varphi(\cdot)$ is the sigmoid function. Hence, we have $0< \Gamma(\mathbf{x}) <10$. This is repeated for all pixels.

\begin{figure*}[t]
    \centering
    \subfigure[]{
    \includegraphics[width=0.19\linewidth]{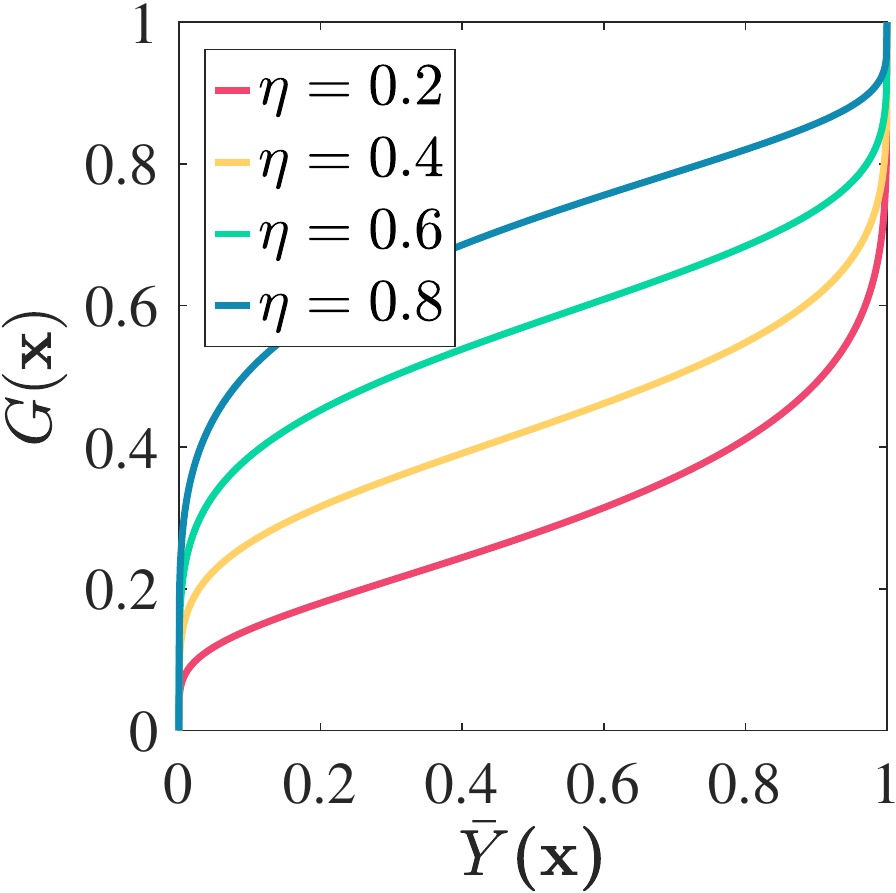}
    }
    \subfigure[$\eta=0.2$]{
    \includegraphics[width=0.185\linewidth]{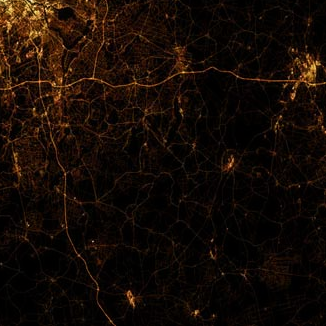}
    }
    \subfigure[$\eta=0.4$]{
    \includegraphics[width=0.185\linewidth]{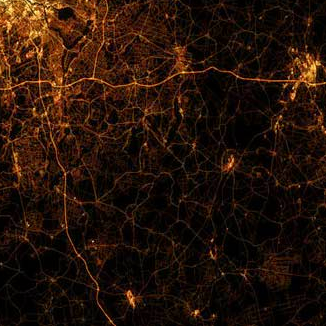}
    }
    \subfigure[$\eta=0.6$]{
    \includegraphics[width=0.185\linewidth]{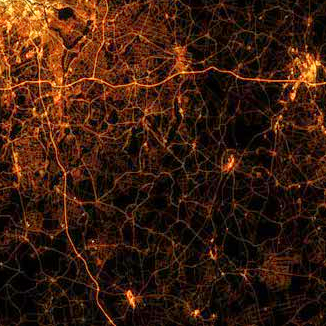}
    }
    \subfigure[$\eta=0.8$]{
    \includegraphics[width=0.185\linewidth]{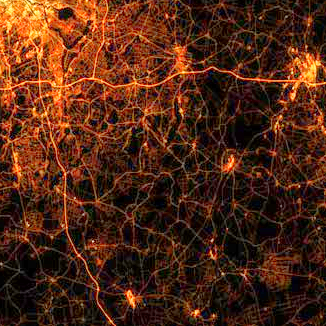}
    }\vspace{-0.1cm}
    \caption{Impacts of the exposure level $\eta$: (a) transformation functions and (b)$\sim$(e) enhanced images $\mathbf{J}$.}
    \label{fig:impact}
\end{figure*}

\subsubsection{Color restoration}
For color restoration, we first obtain the gamma-corrected luminance image $Z$ by transforming each pixel in $Y$ using its gamma value, given by
\begin{equation}
Z(\mathbf{x})=Y_{\max}\left(\frac{Y(\mathbf{x})}{Y_{\max}}\right)^{\Gamma(\mathbf{x})}
\label{eq:enhanced_lum}
\end{equation}
where $Y_{\max}$ is the maximum luminance value (typically 255).
Then, to generate the enhanced RGB color image $\mathbf{J}$, we adopt the simple color restoration approach \cite{schlick1994quantization},
\begin{equation}
\mathbf{J(\mathbf{x})}= \frac{Z(\mathbf{x})}{Y(\mathbf{x})} \mathbf{I(x)},
\end{equation}
which preserves the color ratios.

\subsubsection{Personalized initial $\eta$}
For user's convenience, we provide an initially enhanced image and allow the user to further enhance it interactively. To this end, we estimate an initial exposure level $\eta_{\textrm{init}}$ using a quadratic polynomial by
\begin{equation}
\eta_{\textrm{init}} = a y^2 + b y + c
\label{eq:init}
\end{equation}
where $y=\frac{1}{N}\sum_{\mathbf{x}} Y(\mathbf{x})$ is the average luminance of the input image and $N$ is the number of pixels. The coefficients $a$, $b$, and $c$ are obtained from the observations $\{ (y_i, \eta_i) \}^{M}_{i=1}$ using the method of least squares, where $\eta_i$ is the exposure level selected by the user to enhance the $i$th input image with the average luminance $y_i$. We perform this personalization when $M>3$.

\subsection{Loss Function}

To train IceNet, we use three differentiable losses: interactive brightness control loss $\mathcal{L}_{\textrm{ibc}}$, entropy loss $\mathcal{L}_{\textrm{ent}}$, and smoothness loss $\mathcal{L}_{\textrm{smo}}$. Let us describe these losses subsequently.

\subsubsection{Interactive brightness control loss}
Given an exposure level $\eta$ and a scribble map $S$, we construct a target brightness map $T$ and define the interactive brightness control loss $\mathcal{L}_{\textrm{ibc}}$ as
\begin{equation}
\mathcal{L}_{\textrm{ibc}} = \frac{1}{N}\sum_{\mathbf{x}}(Z(\mathbf{x})-T(\mathbf{x}))^{2}
\label{eq:ibc}
\end{equation}
so that the enhanced luminance image $Z$ in Eq.~\eqref{eq:enhanced_lum} suits the user preference.

To obtain the target brightness $T$, we first add the scribble $S$ to the input luminance $Y$,
\begin{equation}
\tilde{Y} = Y+\lambda S
\label{eq:add}
\end{equation}
where $\lambda$ is a parameter controlling the impacts of scribbles, which is fixed to $5$. By adding $S$, we control local brightness levels. Next, we normalize $\tilde{Y}$ to the range of $[0, 1]$, yielding $\bar{Y}$. Then, we use the bilateral gamma adjustment scheme \cite{lee2007efficient} to improve the visibility of details in dark and bright regions, which performs
\begin{eqnarray}
G_{\textrm{dark}} &=& \bar{Y}^{1/\gamma},\label{eq:Ga}\\
G_{\textrm{bright}} &=& 1-(1-\bar{Y})^{1/\gamma},\label{eq:Gb}\\
G &=& \eta G_{\textrm{dark}} + (1-\eta) G_{\textrm{bright}} \label{eq:G},
\end{eqnarray}
where $\gamma=5$. Note that dark and bright regions are enhanced mainly via Eq.~(\ref{eq:Ga}) and (\ref{eq:Gb}), respectively. To preserve the details in both dark and bright regions, Eq.~(\ref{eq:G}) combines the two results. The user can control the overall brightness of $G$ with the exposure level $\eta$. Figure \ref{fig:impact} illustrates the impacts of $\eta$. Figure \ref{fig:impact}(a) shows the transformation functions from $\bar{Y}(\mathbf{x})$ to $G(\mathbf{x})$, when $\alpha = 0.2, 0.4, 0.6,$ and $0.8$. \modify{The proposed transformation functions have steep slopes near the minimum and maximum levels, so they enable IceNet to enhance both under- and over-exposed regions effectively.} Figure \ref{fig:impact}(b)$\sim$(e) are examples of enhanced images. We see that, as $\eta$ gets bigger, the enhanced image $\mathbf{J}$ becomes brighter.

\begin{figure*}[t]
    \centering
    \subfigure[]{
    \includegraphics[width=0.21\linewidth]{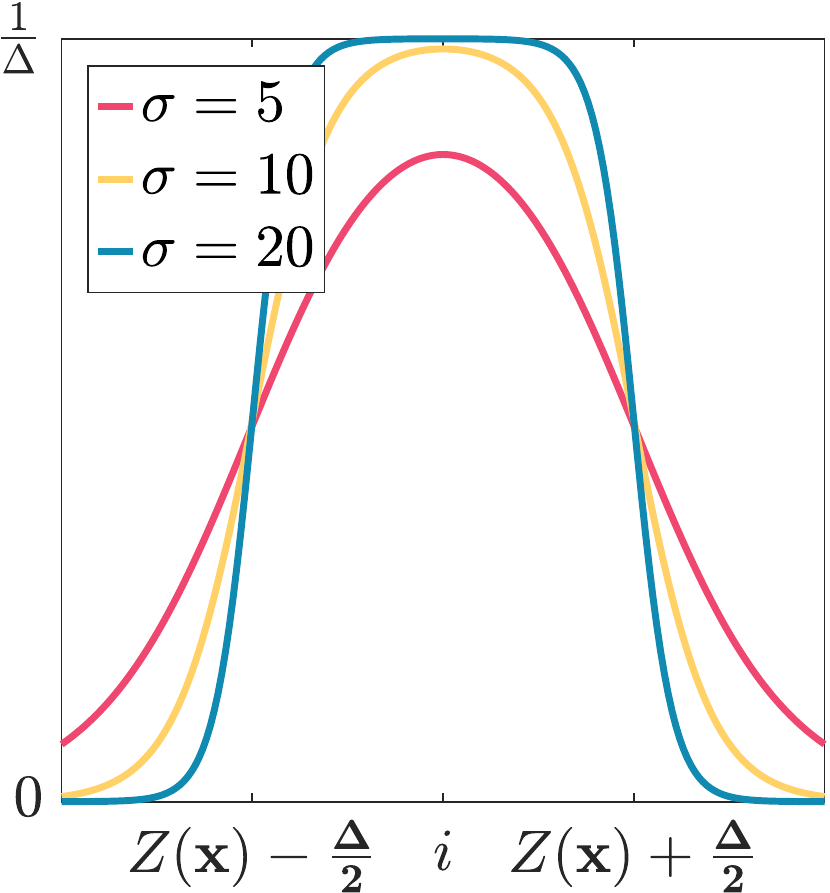}
    }
    \subfigure[IceNet]{
    \includegraphics[width=0.18\linewidth]{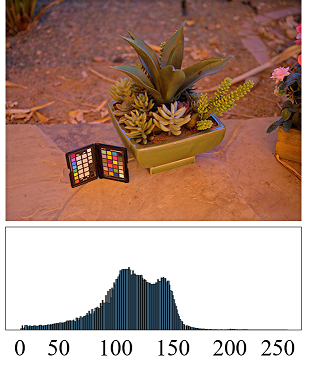}
    }
    \subfigure[w/o $\mathcal{L}_{\textrm{ibc}}$]{
    \includegraphics[width=0.18\linewidth]{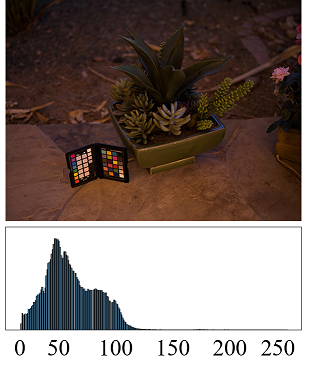}
    }
    \subfigure[w/o $\mathcal{L}_{\textrm{ent}}$]{
    \includegraphics[width=0.18\linewidth]{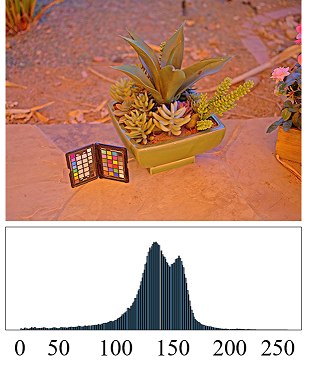}
    }
    \subfigure[w/o $\mathcal{L}_{\textrm{smo}}$]{
    \includegraphics[width=0.18\linewidth]{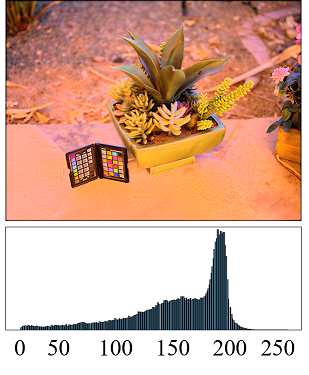}
    }\vspace{-0.1cm}
    \caption{(a) Soft mapping functions. (b)$\sim$(e) Comparison of results of IceNet trained with different combinations of the three loss terms. The corresponding histograms are at the bottom row.}
    \label{fig:entropy}
    %\vspace*{0.4cm}
\end{figure*}

Next, for each pixel $\mathbf{x}$, we compute a local maximum value and scale it by
\begin{equation}
T(\mathbf{x}) = Y_{\max} \times \max_{\mathbf{y}\in\Omega(\mathbf{x})} G(\mathbf{y})
\end{equation}
where $\Omega(\mathbf{x})$ is the $15 \times 15$ window around $\mathbf{x}$. In Eq.~\eqref{eq:ibc}, we use $T$ instead of $Y_{\max}\times G$ to suppress too different target values between adjacent pixels and achieve more reliable enhancement.

\subsubsection{Entropy loss}
An image has a bigger entropy and conveys more information when its pixel values are more widely distributed \cite{cover1999elements}. In particular, the maximum entropy is achieved by the uniform distribution. Thus, we adopt the entropy loss to increase the global contrast by equalizing the histogram of an output image. However, the histogram cannot be directly used because of its non-differentiability. To overcome this issue, we design a soft-histogram. While one pixel contributes to only a single bin of the ordinary histogram, it does to multiple bins in the soft-histogram in a differentiable manner.

For the soft-histogram, we represent the contribution of pixel $\mathbf{x}$ to the $i$th bin using the mapping function in \cite{alivanoglou2008probabilistic}, given by
\begin{equation}
\kappa(\mathbf{x}, i) = \frac{1}{\Delta}\bigg(\varphi\Big(\sigma\big(i-Z(\mathbf{x})+\frac{\Delta}{2}\big)\Big) - \varphi\Big(\sigma\big(i-Z(\mathbf{x})-\frac{\Delta}{2}\big)\Big)\bigg) \\
\end{equation}
where $\varphi(\cdot)$ is the sigmoid function, $\Delta=1$ is the bin width, and $\sigma$ is the hyper-parameter for the slope. Figure \ref{fig:entropy}(a) shows the soft mapping functions when $\sigma=5, 10,$ or $20$. \modify{Note that $\sigma$ controls the tradeoff between the width and height of the mapping function. As $\sigma$ increases, the mapping function gets narrower but taller.} In this work, we set $\sigma=10$. Next, we make the soft-histogram $h$ by summing the contributions of all pixels,
\begin{equation}
h(i) = \sum_{\mathbf{x}}{\kappa(\mathbf{x}, i)}.
\end{equation}
Then, we define $\mathcal{L}_{\textrm{ent}}$ as the inverse of the entropy given by
\begin{equation}
\mathcal{L}_{\textrm{ent}}  = \bigg(-\sum_{i=0}^{Y_{\max}} \frac{h(i)}{N} \log \frac{h(i)}{N} \bigg)^{-1}.
\end{equation}

%------------------------------------------------------------------------
\begin{figure}[t!]
\centering
\setlength\tabcolsep{1.8pt}
\footnotesize
    \begin{tabularx}{\linewidth}{ccc}
        \subfigure{\includegraphics[width=0.32\linewidth]{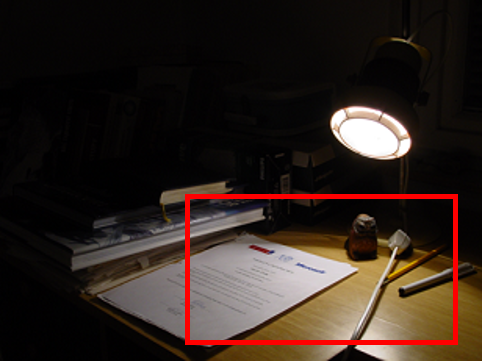}}&
        \subfigure{\includegraphics[width=0.32\linewidth]{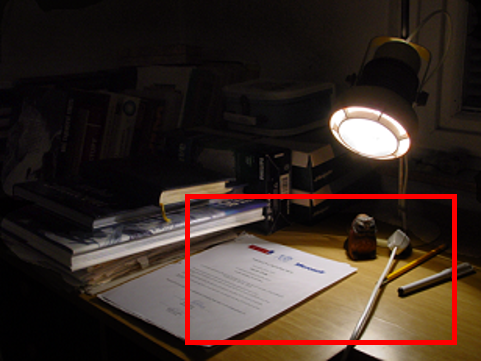}}&
        \subfigure{\includegraphics[width=0.32\linewidth]{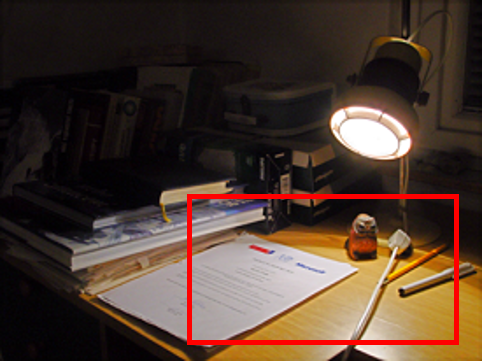}}
    \end{tabularx}
    \begin{tabularx}{\linewidth}{ccc}
        \subfigure{\includegraphics[width=0.32\linewidth]{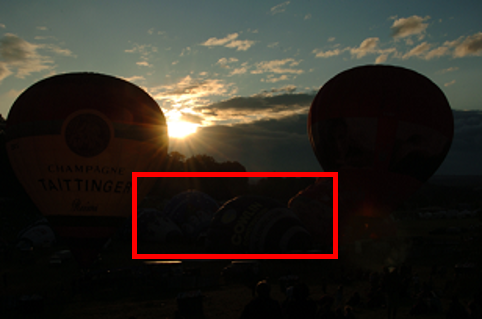}}&
        \subfigure{\includegraphics[width=0.32\linewidth]{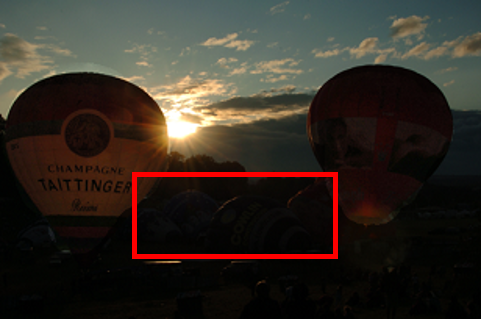}}&
        \subfigure{\includegraphics[width=0.32\linewidth]{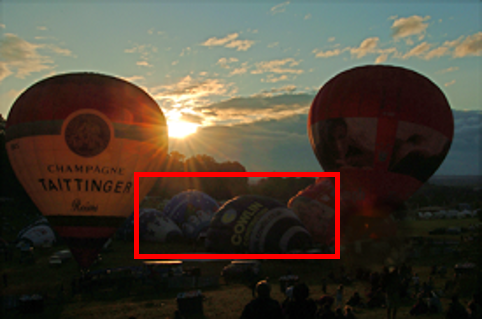}}
    \end{tabularx}
    \begin{tabularx}{\linewidth}{ccc}
        \subfigure{\includegraphics[width=0.32\linewidth]{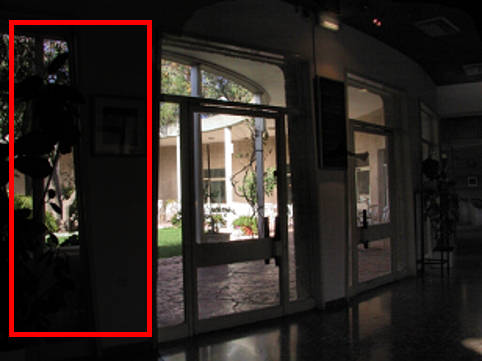}}&
        \subfigure{\includegraphics[width=0.32\linewidth]{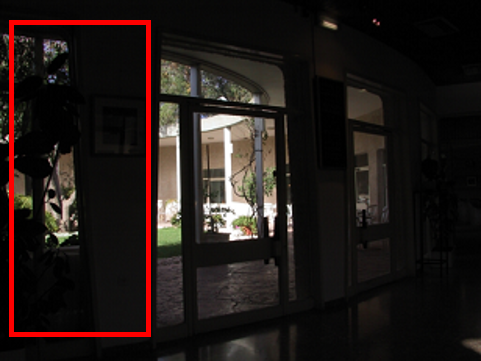}}&
        \subfigure{\includegraphics[width=0.32\linewidth]{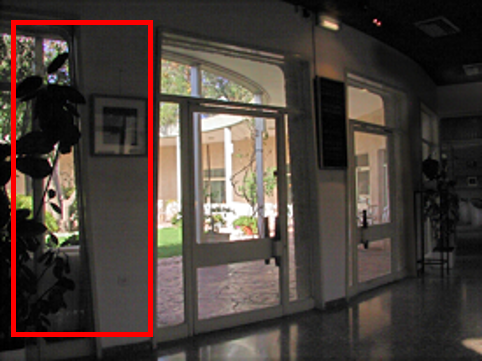}}
    \end{tabularx}
    \begin{tabularx}{\linewidth}{ccc}
        \subfigure{\includegraphics[width=0.32\linewidth]{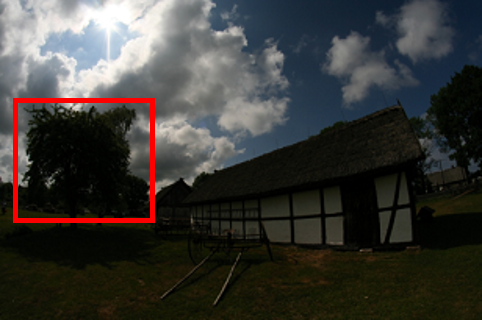}}&
        \subfigure{\includegraphics[width=0.32\linewidth]{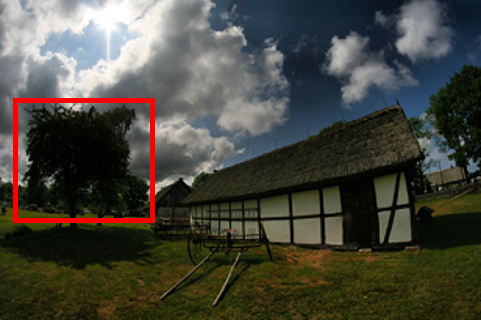}}&
        \subfigure{\includegraphics[width=0.32\linewidth]{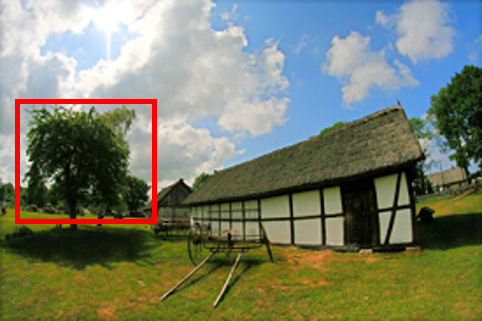}}\\
        Input image & CB \cite{dodgson2009contrast} & Proposed IceNet
    \end{tabularx}\vspace{0.1cm}
    \caption{Qualitative comparison of IceNet with the conventional interactive algorithm
    \cite{dodgson2009contrast}. \modify{Especially, within the red rectangles, the conventional algorithm fails to improve the contrast.}}
    \label{fig:results_inter}
\end{figure}
%------------------------------------------------------------------------

\subsubsection{Smoothness loss}
To encourage smooth variations between neighboring values in the gamma map $\Gamma$ in Eq.~\eqref{eq:gamma_map}, we introduce the smoothness loss
\begin{equation}
\mathcal{L}_{\textrm{smo}} = \|\nabla_h \Gamma\|_F^2 + \|\nabla_v \Gamma\|_F^2
\end{equation}
where $\|\cdot\|_F$ denotes the Frobenius norm, and $\nabla_h$ and $\nabla_v$ are the horizontal and vertical difference operators, respectively.

%------------------------------------------------------------------------
\begin{figure*}[t!]
\centering
\setlength\tabcolsep{1.8pt}
\footnotesize
    \begin{tabularx}{\linewidth}{cccc}
        \subfigure{\includegraphics[width=0.24\linewidth]{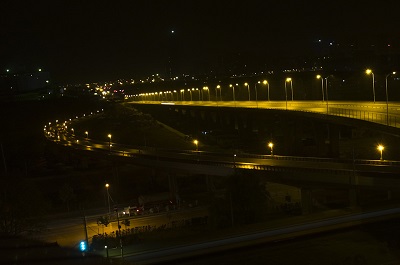}}&
        \subfigure{\includegraphics[width=0.24\linewidth]{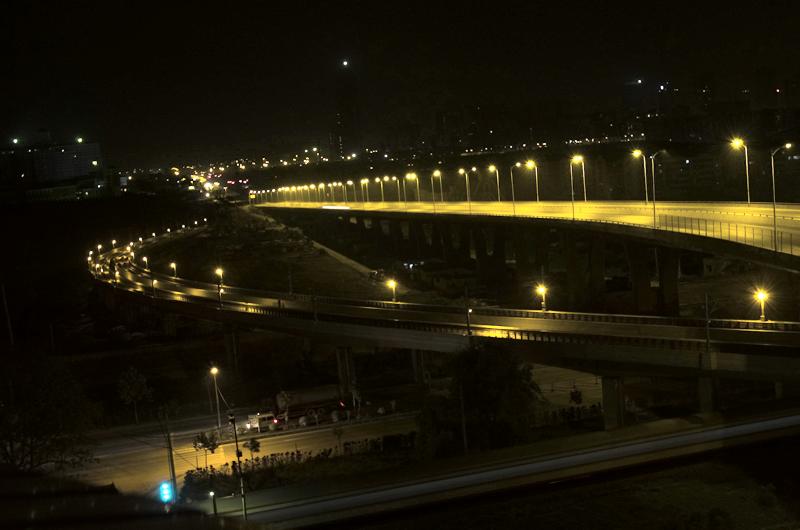}}&
        \subfigure{\includegraphics[width=0.24\linewidth]{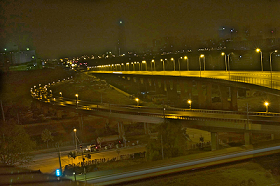}}&
        \subfigure{\includegraphics[width=0.24\linewidth]{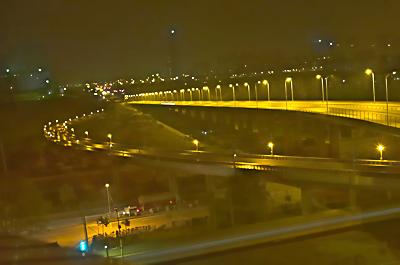}}\\
        Input image & LDR \cite{lee2013contrast} &
        SRIE \cite{fu2016weighted} & Li \etal \cite{li2018structure}
    \end{tabularx}\vspace{0.1cm}
    \begin{tabularx}{\linewidth}{cccc}
        \subfigure{\includegraphics[width=0.24\linewidth]{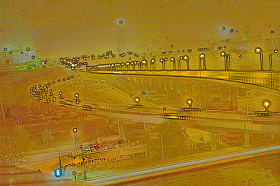}}&
        \subfigure{\includegraphics[width=0.24\linewidth]{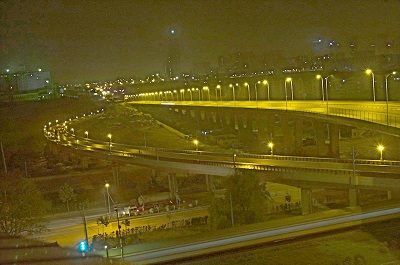}}&
        \subfigure{\includegraphics[width=0.24\linewidth]{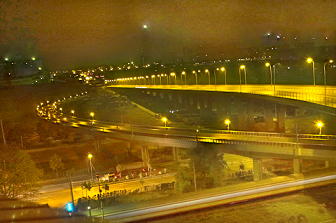}}&
        \subfigure{\includegraphics[width=0.24\linewidth]{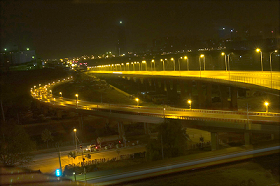}}\\
        RetinexNet \cite{wei2018deep} & Zero-DCE \cite{guo2020zero} &
        EnlightenGAN \cite{jiang2021enlightengan} & Proposed IceNet
    \end{tabularx}\vspace{0.2cm}
    \begin{tabularx}{\linewidth}{cccc}
        \subfigure{\includegraphics[width=0.24\linewidth]{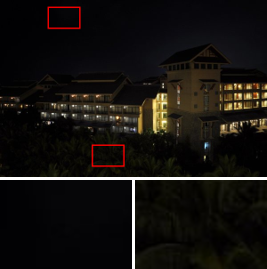}}&
        \subfigure{\includegraphics[width=0.24\linewidth]{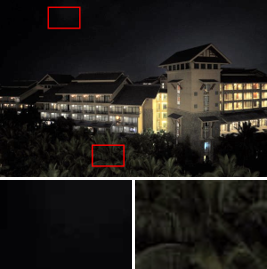}}&
        \subfigure{\includegraphics[width=0.24\linewidth]{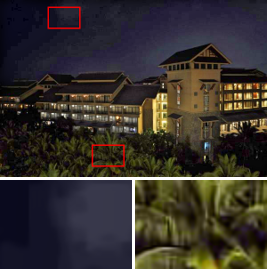}}&
        \subfigure{\includegraphics[width=0.24\linewidth]{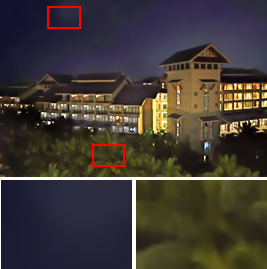}}\\
        Input image & LDR \cite{lee2013contrast} &
        SRIE \cite{fu2016weighted} & Li \etal \cite{li2018structure}
    \end{tabularx}\vspace{0.1cm}
    \begin{tabularx}{\linewidth}{cccc}
        \subfigure{\includegraphics[width=0.24\linewidth]{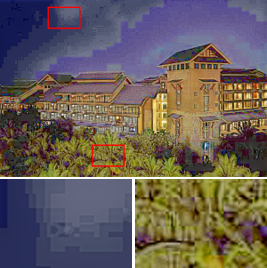}}&
        \subfigure{\includegraphics[width=0.24\linewidth]{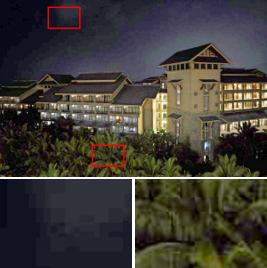}}&
        \subfigure{\includegraphics[width=0.24\linewidth]{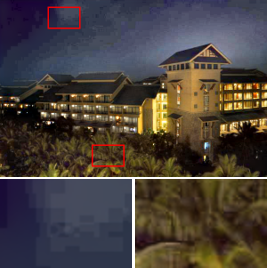}}&
        \subfigure{\includegraphics[width=0.24\linewidth]{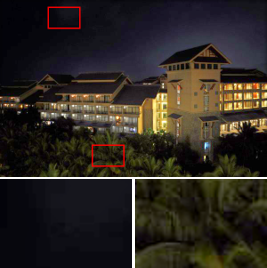}}\\
        RetinexNet \cite{wei2018deep} & Zero-DCE \cite{guo2020zero} &
        EnlightenGAN \cite{jiang2021enlightengan} & Proposed IceNet
    \end{tabularx}\vspace{0.1cm}
    \caption{Qualitative comparison of IceNet with conventional algorithms. Traditional enhancement methods are in the odd rows, while CNN-based methods are in the even rows.}
    \label{fig:qualitative_results}
\end{figure*}
%------------------------------------------------------------------------

\subsubsection{Total loss}
The overall loss is defined as a weighted sum of the three losses, given by
\begin{equation}
\mathcal{L} = \mathcal{L}_{\textrm{ibc}} + w_{\textrm{ent}}\mathcal{L}_{\textrm{ent}} + w_{\textrm{smo}}\mathcal{L}_{\textrm{smo}}
\label{eq:overall_loss}
\end{equation}
where $w_{\textrm{ent}}$ and $w_{\textrm{smo}}$ are weights. \modify{Thus, the proposed IceNet is trained to minimize the three loss functions. First, $\mathcal{L}_{\textrm{ibc}}$ enables IceNet to control global and local brightness. Second, $\mathcal{L}_{\textrm{ent}}$ encourages a flat histogram, which can increase the global contrast. Third, $\mathcal{L}_{\textrm{smo}}$ smooths the gamma map. Figure \ref{fig:entropy}(b)$\sim$(e) illustrate the efficacy of each loss, as will be discussed in Section \ref{sec:ablation_loss}.}

%------------------------------------------------------------------------
\begin{figure*}[t!]
\centering
\setlength\tabcolsep{1.8pt}
\footnotesize
    \begin{tabularx}{\linewidth}{cccc}
        \subfigure{\includegraphics[width=0.24\linewidth]{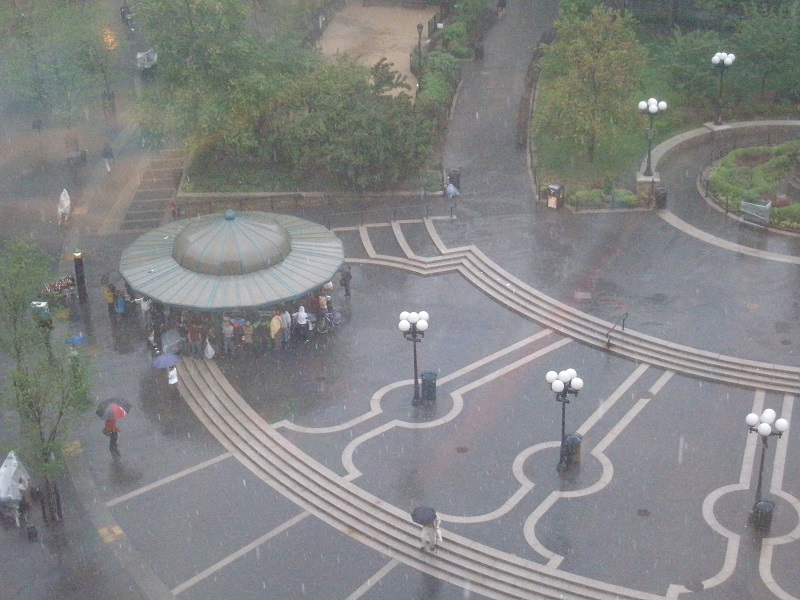}}&
        \subfigure{\includegraphics[width=0.24\linewidth]{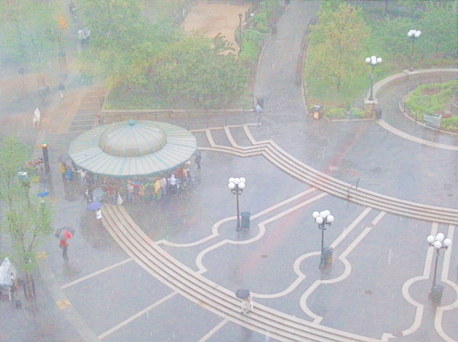}}&
        \subfigure{\includegraphics[width=0.24\linewidth]{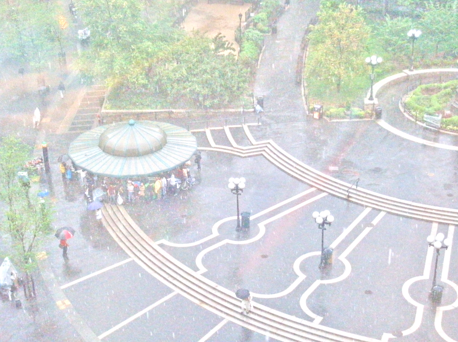}}&
        \subfigure{\includegraphics[width=0.24\linewidth]{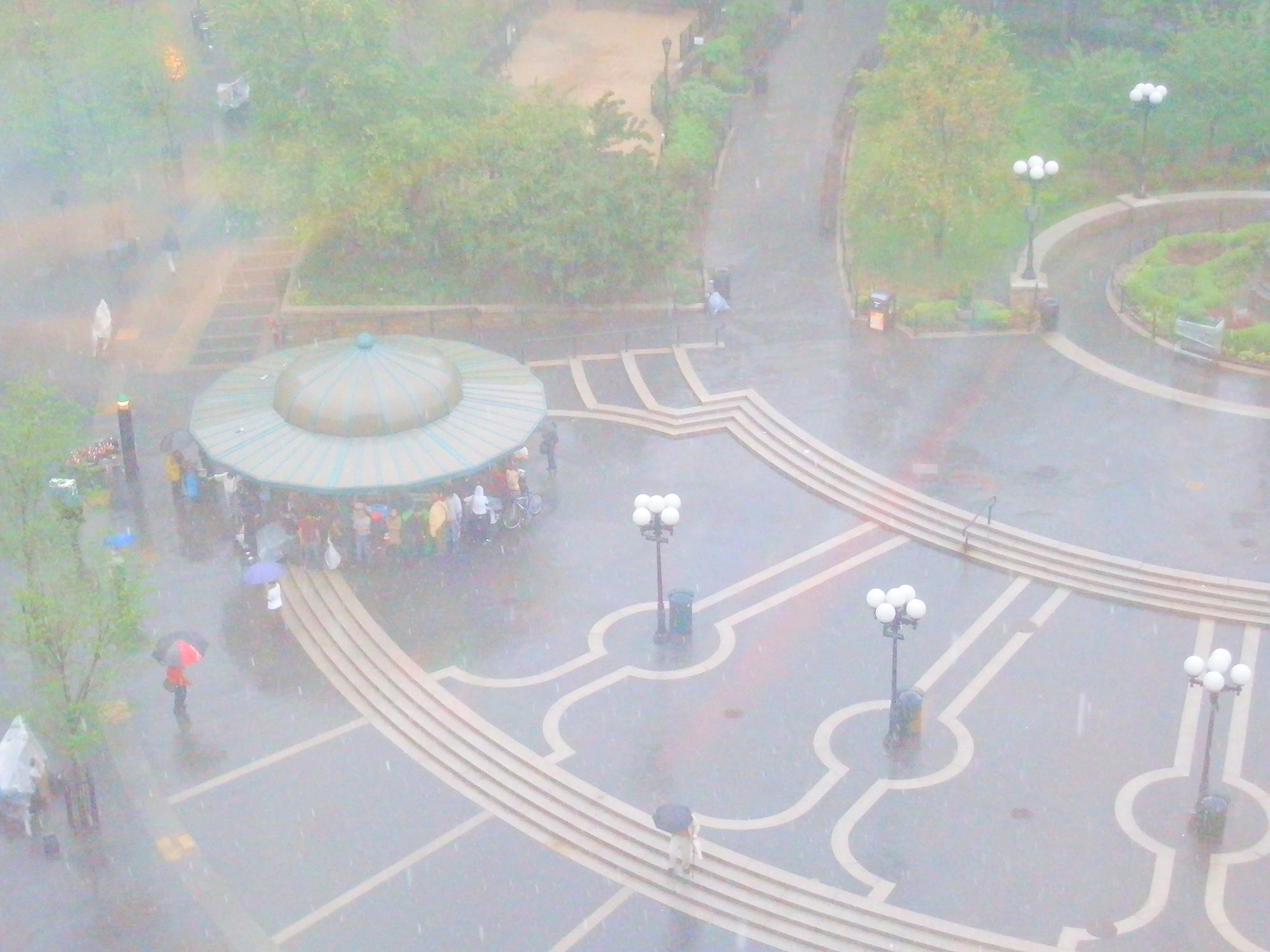}}\\
        Input image & SRIE \cite{fu2016weighted} &
        LIME \cite{guo2017lime} & Li \etal \cite{li2018structure}
    \end{tabularx}\vspace{0.1cm}
    \begin{tabularx}{\linewidth}{cccc}
        \subfigure{\includegraphics[width=0.24\linewidth]{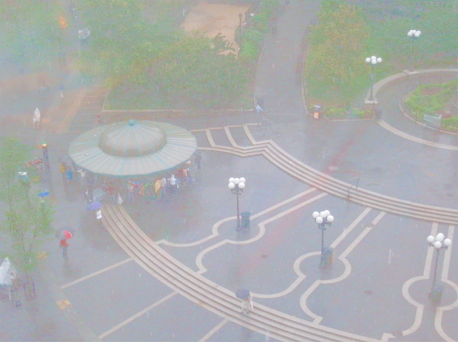}}&
        \subfigure{\includegraphics[width=0.24\linewidth]{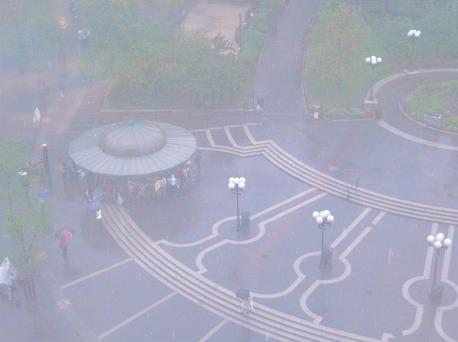}}&
        \subfigure{\includegraphics[width=0.24\linewidth]{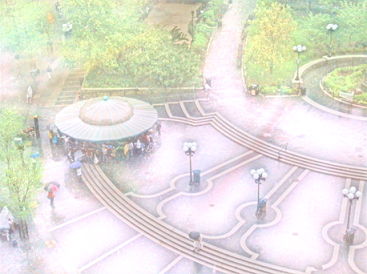}}&
        \subfigure{\includegraphics[width=0.24\linewidth]{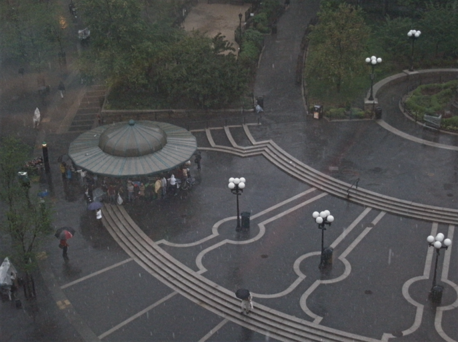}}\\
        RetinexNet \cite{wei2018deep} & Zero-DCE \cite{guo2020zero} &
        EnlightenGAN \cite{jiang2021enlightengan} & Proposed IceNet
    \end{tabularx}\vspace{0.2cm}
    \begin{tabularx}{\linewidth}{cccc}
        \subfigure{\includegraphics[width=0.24\linewidth]{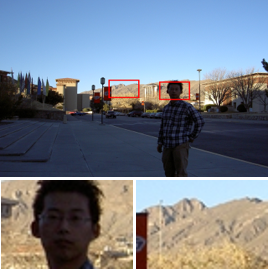}}&
        \subfigure{\includegraphics[width=0.24\linewidth]{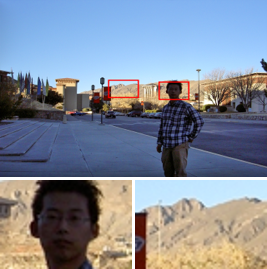}}&
        \subfigure{\includegraphics[width=0.24\linewidth]{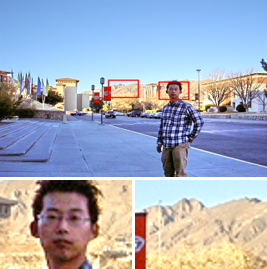}}&
        \subfigure{\includegraphics[width=0.24\linewidth]{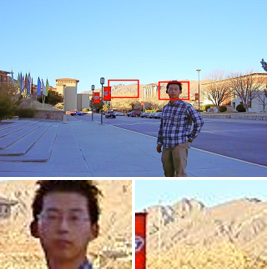}}\\
        Input image & SRIE \cite{fu2016weighted} &
        LIME \cite{guo2017lime} & Li \etal \cite{li2018structure}
    \end{tabularx}\vspace{0.1cm}
    \begin{tabularx}{\linewidth}{cccc}
        \subfigure{\includegraphics[width=0.24\linewidth]{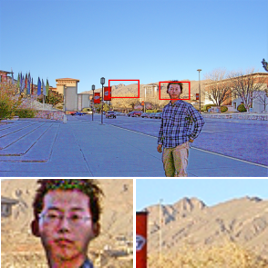}}&
        \subfigure{\includegraphics[width=0.24\linewidth]{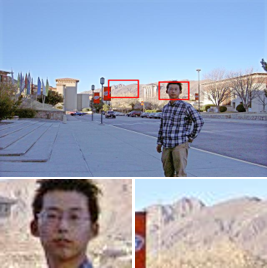}}&
        \subfigure{\includegraphics[width=0.24\linewidth]{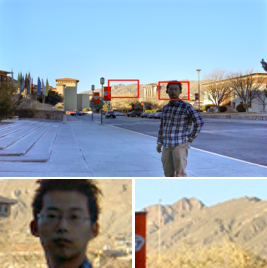}}&
        \subfigure{\includegraphics[width=0.24\linewidth]{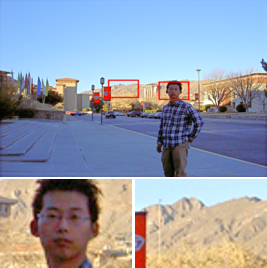}}\\
        RetinexNet \cite{wei2018deep}] & Zero-DCE \cite{guo2020zero} &
        EnlightenGAN \cite{jiang2021enlightengan} & Proposed IceNet
    \end{tabularx}\vspace{0.1cm}
    \caption{Qualitative comparison of IceNet with conventional algorithms. Traditional enhancement methods are in the odd rows, while CNN-based methods are in the even rows.}
    \label{fig:qualitative_results2}
\end{figure*}
%------------------------------------------------------------------------

\subsection{Implementation Details}

The number of output channels of every convolutional or fully-connected layer is 32. In every convolutional layer, the zero padding and the ReLU activation are performed. The batch normalization is not applied, since a small mini-batch size of 8 is used. \modify{We initialize all parameters in IceNet with Gaussian random numbers. Then, we update the parameters via the Adam optimizer \cite{kingma2015adam} with an initial learning rate of $10^{-3}$.} We employ the same 2,002 training images as \cite{guo2020zero}. The training images are randomly selected in the Part1 subset of SICE \cite{cai2018learning}, including under-, normal-, and over-exposed images. The training is iterated for 50 epochs with an RTX 2080Ti GPU, which takes about 30 minutes only. All training images are resized to $512\times512$. To emulate user annotations, an exposure level $\eta$ is randomly selected from $[0.2, 0.8]$ and red and blue scribbles, respectively, are generated 0$\sim$5 times at random positions. For this emulation, each scribble is a circle of a radius of 10 pixels. In Eq.~\eqref{eq:overall_loss}, $w_{\textrm{ent}}=10$~and~$w_{\textrm{smo}}=20$.

\vspace{0.2cm}
\section{Experiments}

We compare the proposed algorithm with an interactive enhancement algorithm (CB \cite{dodgson2009contrast}) and seven conventional ones (LDR \cite{lee2013contrast}, SRIE \cite{fu2016weighted}, LIME \cite{guo2017lime}, Li \etal \cite{li2018structure}, RetinexNet \cite{wei2018deep}, Zero-DCE \cite{guo2020zero}, EnlightenGAN \cite{jiang2021enlightengan}). Next, we conduct an analysis of preference for contrast. \modify{All experiments are conducted in Python with an RTX 2080Ti GPU and a Ryzen 9 3900X CPU.}

\subsection{Comparative Assessment}\label{sec:userstudy}

In this section, we obtain enhanced images of the conventional algorithms using the source codes and parameters provided by the authors, unless otherwise specified.

%------------------------------------------------------------------------
\begin{figure*}[t!]
\centering
\setlength\tabcolsep{1.8pt}
\footnotesize
    \begin{tabularx}{\linewidth}{cccc}
        \subfigure{\includegraphics[width=0.24\linewidth]{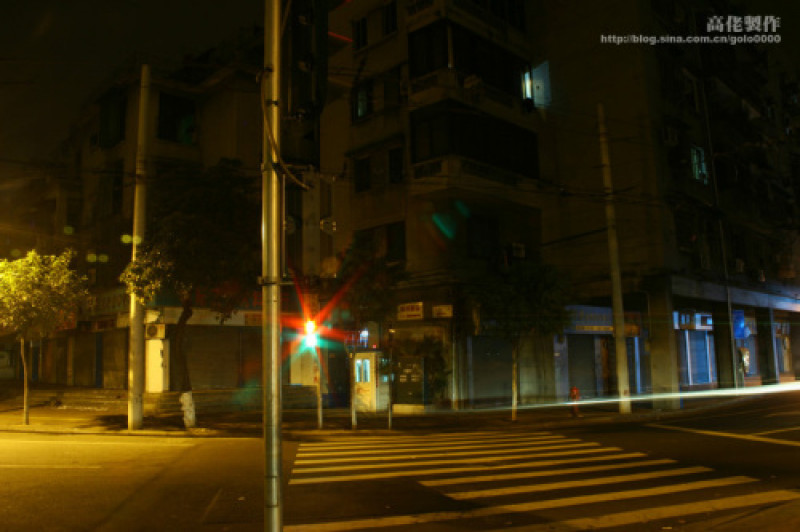}}&
        \subfigure{\includegraphics[width=0.24\linewidth]{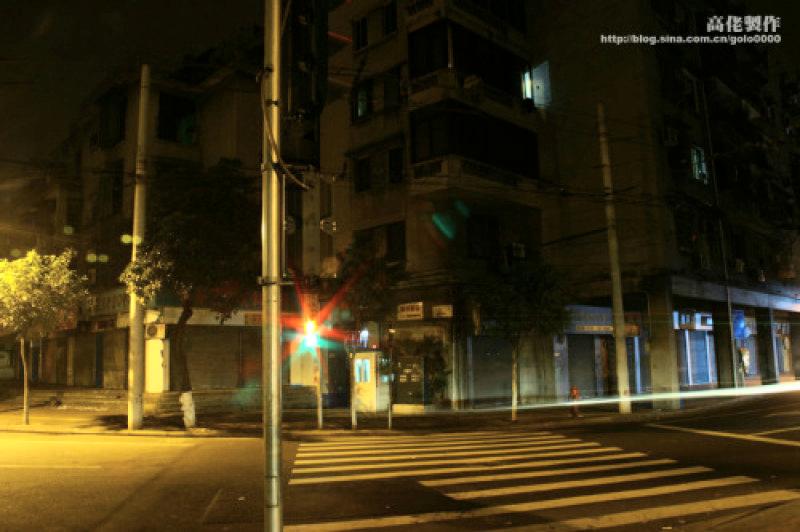}}&
        \subfigure{\includegraphics[width=0.24\linewidth]{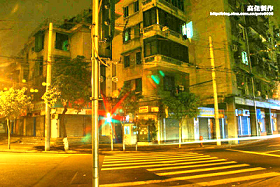}}&
        \subfigure{\includegraphics[width=0.24\linewidth]{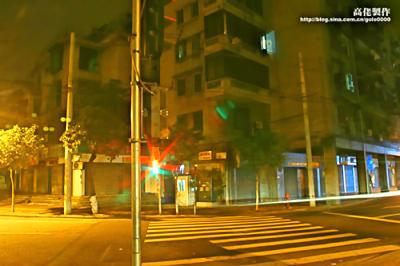}}\\
        Input image & LDR \cite{lee2013contrast} &
        LIME \cite{guo2017lime} & Li \etal \cite{li2018structure}
    \end{tabularx}\vspace{0.1cm}
    \begin{tabularx}{\linewidth}{cccc}
        \subfigure{\includegraphics[width=0.24\linewidth]{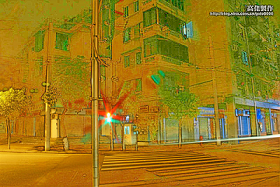}}&
        \subfigure{\includegraphics[width=0.24\linewidth]{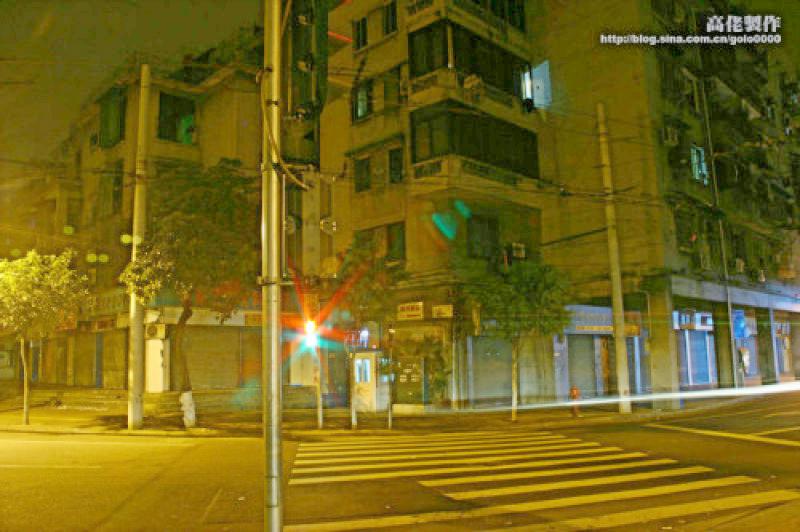}}&
        \subfigure{\includegraphics[width=0.24\linewidth]{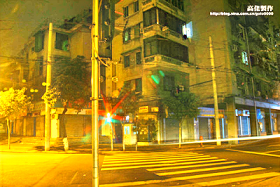}}&
        \subfigure{\includegraphics[width=0.24\linewidth]{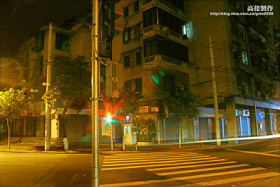}}\\
        RetinexNet \cite{wei2018deep} & Zero-DCE \cite{guo2020zero} &
        EnlightenGAN \cite{jiang2021enlightengan} & Proposed IceNet
    \end{tabularx}\vspace{0.2cm}
    \begin{tabularx}{\linewidth}{cccc}
        \subfigure{\includegraphics[width=0.24\linewidth]{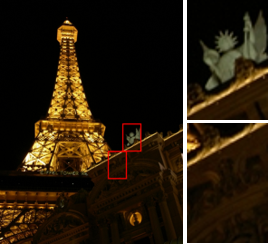}}&
        \subfigure{\includegraphics[width=0.24\linewidth]{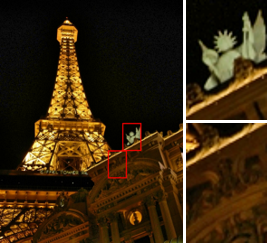}}&
        \subfigure{\includegraphics[width=0.24\linewidth]{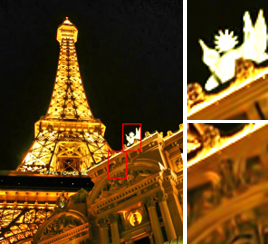}}&
        \subfigure{\includegraphics[width=0.24\linewidth]{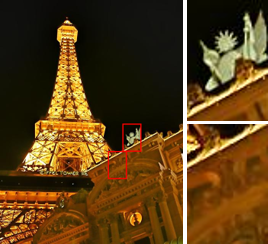}}\\
        Input image & SRIE \cite{fu2016weighted} &
        LIME \cite{guo2017lime} & Li \etal \cite{li2018structure}
    \end{tabularx}\vspace{0.1cm}
    \begin{tabularx}{\linewidth}{cccc}
        \subfigure{\includegraphics[width=0.24\linewidth]{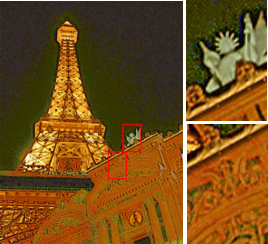}}&
        \subfigure{\includegraphics[width=0.24\linewidth]{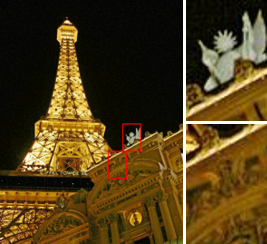}}&
        \subfigure{\includegraphics[width=0.24\linewidth]{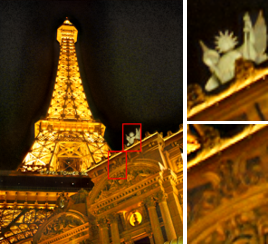}}&
        \subfigure{\includegraphics[width=0.24\linewidth]{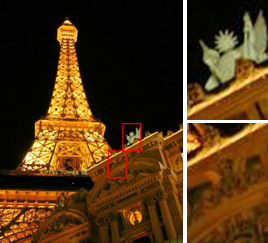}}\\
        RetinexNet \cite{wei2018deep}] & Zero-DCE \cite{guo2020zero} &
        EnlightenGAN \cite{jiang2021enlightengan} & Proposed IceNet
    \end{tabularx}\vspace{0.1cm}
    \caption{\modify{Qualitative comparison of IceNet with conventional algorithms. Traditional enhancement methods are in the odd rows, while CNN-based methods are in the even rows.}}
    \label{fig:qualitative_results3}
\end{figure*}
%------------------------------------------------------------------------

\subsubsection{User studies for IceNet}

Source codes of the interactive methods \cite{grundland2006interactive,lischinski2006interactive, dodgson2009contrast} are not available. Thus, we implemented CB ourselves and compared it with the proposed IceNet on the test set of MEF \cite{ma2015perceptual}. We asked 11 participants to provide annotations to the two interactive methods and vote for better results. 187 votes in total (17 images $\times$ 11 participants) were cast for the preferred interactive methods. The proposed IceNet won significantly more votes: IceNet was preferred in $80.2\%$ of the tests, while CB was in only $19.8\%$. Figure \ref{fig:results_inter} compares qualitative results, in which the results were obtained by the same participant. While CB does not improve the contrast sufficiently, IceNet does it successfully.

%------------------------------------------------------------------------
\begin{figure*}[t!]
\centering
\setlength\tabcolsep{1.8pt}
\footnotesize
    \begin{tabularx}{\linewidth}{cccc}
        \subfigure{\includegraphics[width=0.24\linewidth]{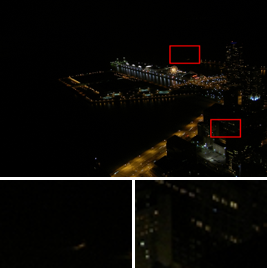}}&
        \subfigure{\includegraphics[width=0.24\linewidth]{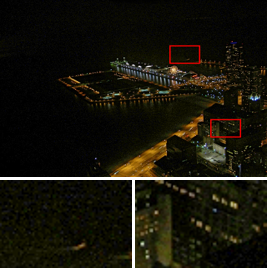}}&
        \subfigure{\includegraphics[width=0.24\linewidth]{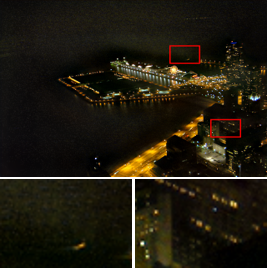}}&
        \subfigure{\includegraphics[width=0.24\linewidth]{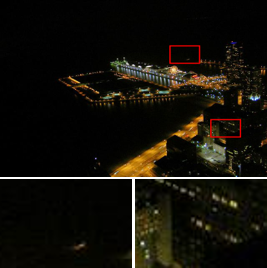}}\\
    \end{tabularx}
    \begin{tabularx}{\linewidth}{cccc}
        \subfigure{\includegraphics[width=0.24\linewidth]{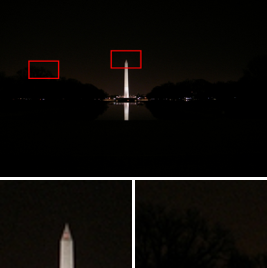}}&
        \subfigure{\includegraphics[width=0.24\linewidth]{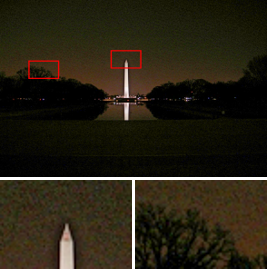}}&
        \subfigure{\includegraphics[width=0.24\linewidth]{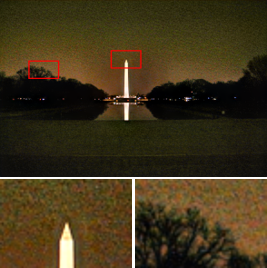}}&
        \subfigure{\includegraphics[width=0.24\linewidth]{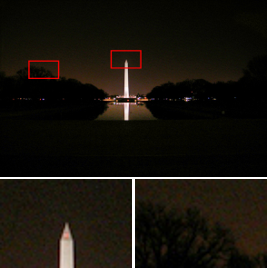}}\\
    \end{tabularx}
    \begin{tabularx}{\linewidth}{cccc}
        \subfigure{\includegraphics[width=0.24\linewidth]{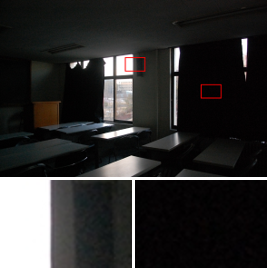}}&
        \subfigure{\includegraphics[width=0.24\linewidth]{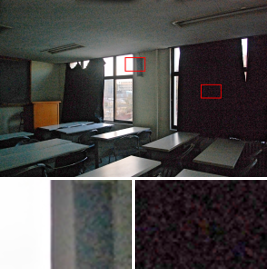}}&
        \subfigure{\includegraphics[width=0.24\linewidth]{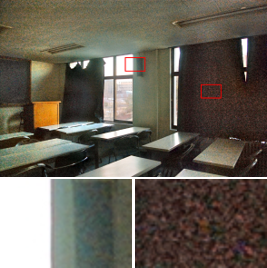}}&
        \subfigure{\includegraphics[width=0.24\linewidth]{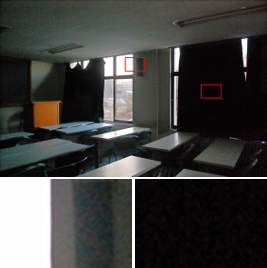}}\\
        Input image & SRIE \cite{fu2016weighted} & EnlightenGAN \cite{jiang2021enlightengan}
        & Proposed IceNet
    \end{tabularx}\vspace{0.1cm}
    \caption{Qualitative comparison of IceNet with conventional algorithms. Note that all enhanced images are obtained automatically without per-image interaction.}%\vspace{0.3cm}
    \label{fig:qualitative_auto}
\end{figure*}
%------------------------------------------------------------------------

%------------------------------------------------------------------------
\begin{table*}[!ht]
    \caption
    {
        User study results. 550 votes in total (50 images $\times$ 11 participants)  were cast for the most preferred methods. \modify{IceNet won the most votes of 314.  It was the most preferred in 57\%, while the second-best SRIE was in only 11\%.}
    }\vspace{-0.1cm}
    \centering
    \resizebox{0.975\linewidth}{!}{
    {\scriptsize
    \begin{tabularx}{\linewidth}{c | C C C C C C C C}
        \toprule
        & LDR & SRIE & LIME & Li \etal & RetinexNet & Zero-DCE & EnlightenGAN & Proposed \\
        & \cite{lee2013contrast} & \cite{fu2016weighted} & \cite{guo2017lime} & \cite{li2018structure} & \cite{wei2018deep} & \cite{guo2020zero} & \cite{jiang2021enlightengan} &  IceNet \\
        \midrule
        \# votes & 37 & 61 & 28 & 35 & 0 & 18 & 57 & 314 \\
        \bottomrule
    \end{tabularx}}}
    \label{table:US}\vspace{0.1cm}
    %\vspace*{0.6cm}
\end{table*}
%------------------------------------------------------------------------

%------------------------------------------------------------------------
\begin{table}[t]
    \caption
    {
        Two user studies for personalized initial $\eta$. (a) Satisfaction survey results for personalized initial $\eta$. (b) 220 votes in total (20 images $\times$ 11 participants) were cast for the most preferred method.
    }\vspace{-0.2cm}
    \centering
    \begin{tabularx}{0.34\linewidth}{cc}
        \multicolumn{2}{c}{\footnotesize{(a)}}\\ \vspace{-0.2cm} \\
        \toprule
        Satisfaction & MAE\\
        \midrule
        $82.5\%$ & 0.068\\
        \bottomrule
    \end{tabularx}
    \hspace{0.4cm}
    \begin{tabularx}{0.52\linewidth}{cCCC}
        \multicolumn{4}{c}{\footnotesize{(b)}}\\ \vspace{-0.2cm} \\
        \toprule
        & \cite{fu2016weighted} & \cite{jiang2021enlightengan} & IceNet\\
        \midrule
        \# votes & 62 & 49 & 109 \\
        \bottomrule
    \end{tabularx}
    \label{US:personal}
\end{table}
%------------------------------------------------------------------------
For more subjective assessment, we collected 50 images by choosing the first 10 indexed images from each of the test sets of NPE \cite{wang2013naturalness} (85 images), LIME \cite{guo2017lime} (10 images), MEF \cite{ma2015perceptual} (17 images), DICM \cite{lee2012contrast, lee2013contrast} (69 images), and VV \cite{vv} (23 images). Then, we conducted another user study with the 11 participants. It was designed as follows:
\begin{enumerate}
\itemsep0mm
\item A participant provides annotations to IceNet, which then yields an enhanced image.
\item The eight enhanced images obtained by IceNet and the seven conventional algorithms are presented to the participant in a random order.
\item The participant votes for the most pleasing result.
\end{enumerate}
It is recommended to watch the supplemental video for a demo of this user study. Note that a participant may prefer an automatically enhanced image of a conventional algorithm to the result of IceNet. This study was conducted to check whether the participants were sufficiently satisfied with their interactive enhancement results using IceNet. Table~\ref{table:US} summarizes the results. The proposed IceNet won the most votes; it was the most preferred in 57\% of the tests, while the second-best SRIE was in only 11\%.

\modify{Figures \ref{fig:qualitative_results}, \ref{fig:qualitative_results2}, and \ref{fig:qualitative_results3} compare qualitative results, in which the results of IceNet were obtained by a participant. IceNet provides more pleasing results with fewer artifacts than the conventional algorithms do. Note that LDR and SRIE do not improve the contrast sufficiently, and Li \etal~yield blurred images. The other conventional algorithms tend to generate noise or do over-enhancement with contour artifacts.}

%------------------------------------------------------------------------
\begin{figure*}[t]
\centering
\setlength\tabcolsep{1.8pt}
\footnotesize
    \begin{tabularx}{\linewidth}{cccc}
        \includegraphics[width=0.24\linewidth]{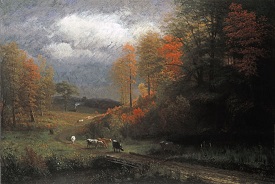}
        &\includegraphics[width=0.24\linewidth]{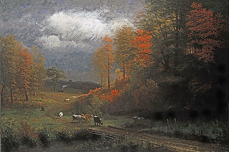}
        &\includegraphics[width=0.24\linewidth]{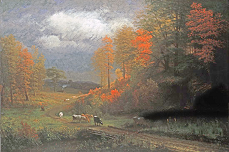}
        &\includegraphics[width=0.24\linewidth]{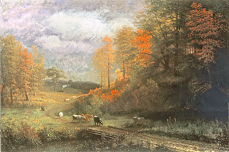}\\
        \multirow{-10}{*}{
            \footnotesize
            \setlength\tabcolsep{4pt}
            \centering
            \begin{tabularx}{0.22\linewidth}{c}
                Input image\\ \\
                \includegraphics[width=0.2\linewidth]{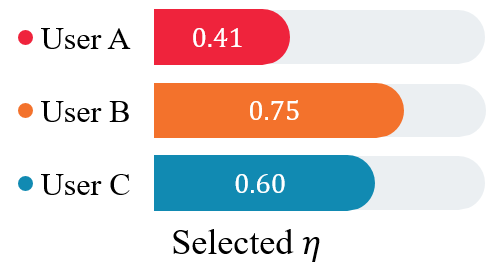}
            \end{tabularx}
        }
        &\includegraphics[width=0.24\linewidth]{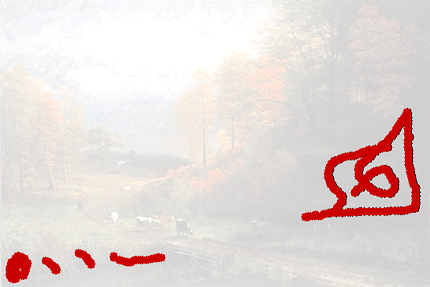}
        &\includegraphics[width=0.24\linewidth]{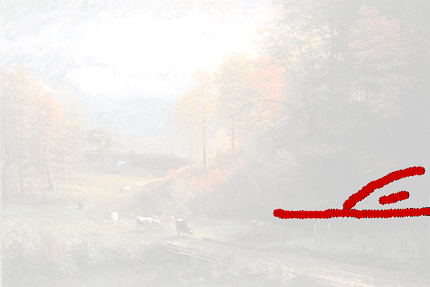}
        &\includegraphics[width=0.24\linewidth]{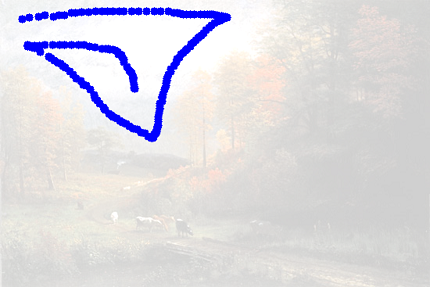}\\ \vspace{-0.2cm} \\
    \end{tabularx}
    %%%%%%%%%%%%%%%%%%%%%%%%%%%%%%%%%%%%%%%%%%%%%%%%%%%%%%%%%%%%%%%%%%%%%%%%%%%
    \begin{tabularx}{\linewidth}{cccc}
        \includegraphics[width=0.24\linewidth]{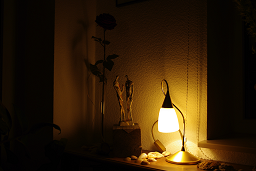}
        &\includegraphics[width=0.24\linewidth]{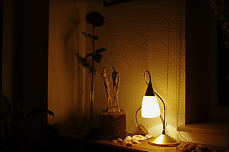}
        &\includegraphics[width=0.24\linewidth]{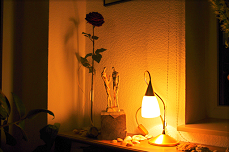}
        &\includegraphics[width=0.24\linewidth]{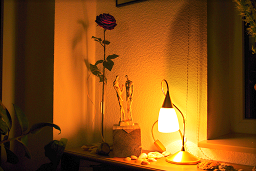}\\
        \multirow{-9.5}{*}{
            \footnotesize
            \setlength\tabcolsep{4pt}
            \centering
            \begin{tabularx}{0.22\linewidth}{c}
                Input image\\ \vspace{0.1cm} \\                \includegraphics[width=0.2\linewidth]{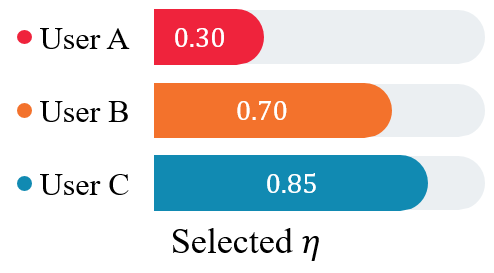}
            \end{tabularx}
        }
        &\includegraphics[width=0.24\linewidth]{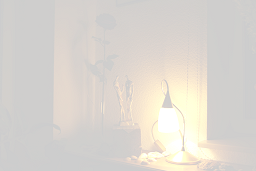}
        &\includegraphics[width=0.24\linewidth]{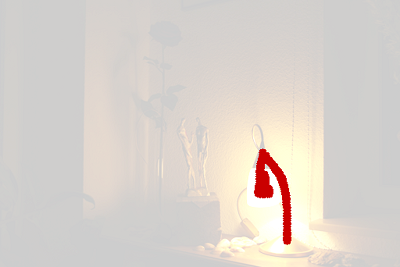}
        &\includegraphics[width=0.24\linewidth]{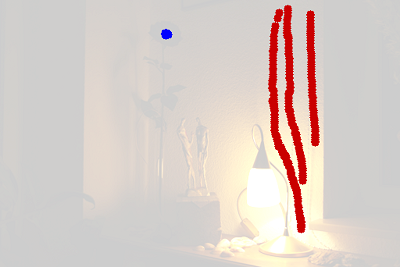}\\
        %%%%%%%%%%%%%%%%%%%%%%%%%%%%%%%%%%%%%%%%%%%%%%%%%%%%%%%%%%%%%%%%%%%%%%%%%%%
         & User A & User B & User C \\
    \end{tabularx}
    \caption{The odd rows show that three participants enhanced the same input image differently, which indicates that people have diverse preferences for contrast. The even rows are the annotations provided by the participants.}
    \vspace{0.05cm}
    \label{fig:userstudy}
\end{figure*}
%------------------------------------------------------------------------
%------------------------------------------------------------------------
\begin{table*}[t]
    \renewcommand{\arraystretch}{1.1}
    \caption
    {
        Full-reference image quality assessment on the Part2 subset of SICE \cite{cai2018learning}. \modify{The best results are boldfaced, and the second-best ones are underlined.}
    }\vspace{-0.1cm}
    \centering
    \resizebox{0.975\linewidth}{!}{
    {\scriptsize
    \begin{tabularx}{\linewidth}{c | C C C C C C C C}
        \toprule
        & LDR & LIME & Li \etal & RetinexNet & Zero-DCE & EnlightenGAN & Proposed & Proposed\\
        & \cite{lee2013contrast} & \cite{guo2017lime} & \cite{li2018structure} & \cite{wei2018deep} & \cite{guo2020zero} & \cite{jiang2021enlightengan} & IceNet & $\text{IceNet}^*$ \\
        \midrule
        PSNR & 13.39 & 16.17 & 15.19 & 15.99 & 16.57 & 16.21 & \underline{17.33} & \bf{19.07} \\
        SSIM & 0.47 & 0.57 & 0.54 & 0.53 & 0.59 & 0.59 & \underline{0.67} & \bf{0.70} \\
        \textcolor{myblue}{Run time (ms)} & \textcolor{myblue}{49} & \textcolor{myblue}{245} & \textcolor{myblue}{52,564} & \textcolor{myblue}{1,023} & \textcolor{myblue}{\bf{2}} & \textcolor{myblue}{\underline{8}} & \textcolor{myblue}{\bf{2}} & \textcolor{myblue}{\bf{2}} \\
        \bottomrule
    \end{tabularx}}}
    \vspace{0.15cm}
    \label{table:QE}
\end{table*}
%------------------------------------------------------------------------

\subsubsection{User studies for personalized initial $\eta$} We asked the 11 participants how satisfied they were with initially enhanced images. For this test, the coefficients in Eq.~\eqref{eq:init} were personalized by the pairs of $(y_i, \eta_i)$, which had been selected by each participant during the previous user study. Then, we collected 20 new test images, by selecting five low contrast images from each of NPE, MEF, DICM, and VV. These new images did not overlap with the images used for the test in Table \ref{table:US}. On average, each user was satisfied by 16.5 initial images out of those 20 cases. In the other word, 82.5 percent of initially enhanced images satisfied the users. Also, the mean absolute error (MAE) between the personalized initial $\eta$ and the fine-tuned one was only 0.068. Table~\ref{US:personal}(a) shows the results in this test. These results confirm that initially enhanced images are satisfactory in most cases.

%------------------------------------------------------------------------
\begin{figure*}[t]
\centering
\setlength\tabcolsep{1.8pt}
\footnotesize
    \begin{tabularx}{\linewidth}{ccccc}
        \subfigure{\includegraphics[width=0.19\linewidth]{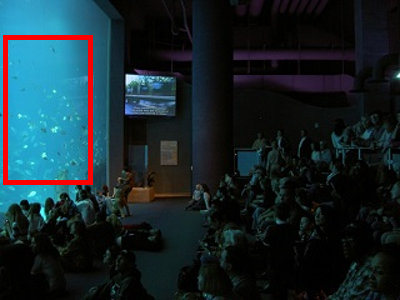}}&
        \subfigure{\includegraphics[width=0.19\linewidth]{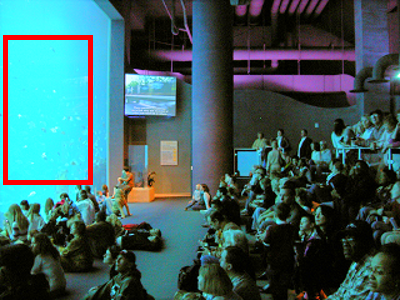}}&
        \subfigure{\includegraphics[width=0.19\linewidth]{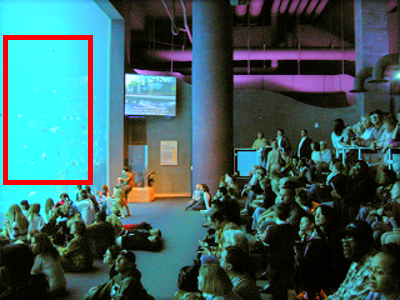}}&
        \subfigure{\includegraphics[width=0.19\linewidth]{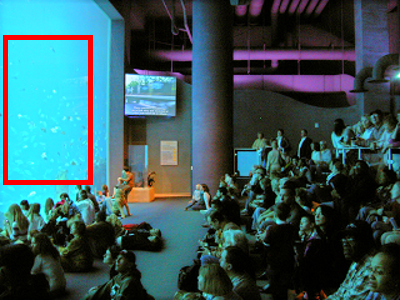}}&
        \subfigure{\includegraphics[width=0.19\linewidth]{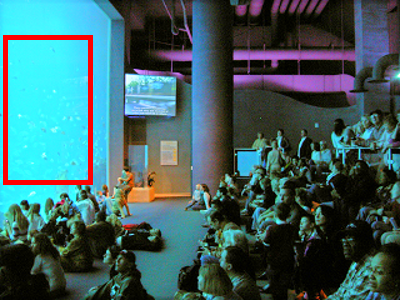}}\\
        \subfigure{\includegraphics[width=0.19\linewidth]{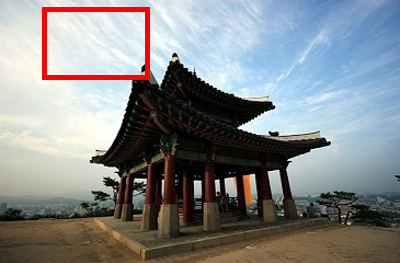}}&
        \subfigure{\includegraphics[width=0.19\linewidth]{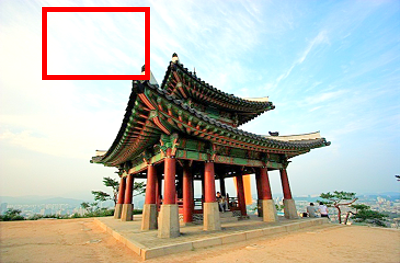}}&
        \subfigure{\includegraphics[width=0.19\linewidth]{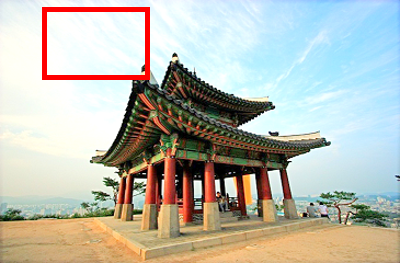}}&
        \subfigure{\includegraphics[width=0.19\linewidth]{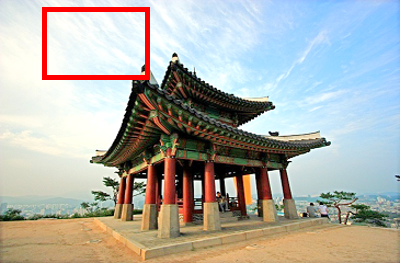}}&
        \subfigure{\includegraphics[width=0.19\linewidth]{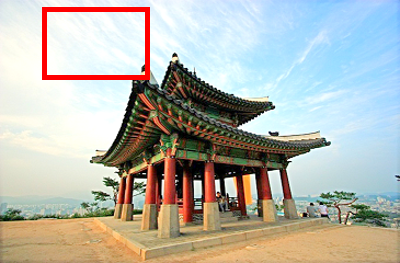}}\\
        Input image & (3, 32, 29K) & (7, 16, 21K) & (7, 32, 85K) & (7, 64, 338K)\\
    \end{tabularx}
\caption{\modify{Impacts of parameter settings. Here, ($L$-$C$-$N$) represents IceNet with $L$ convolutional layers, $C$ output channels, and $N$ parameters. Note that the default setting is (7, 32, 85K). The results are obtained at $\eta=0.95$ without scribbles. The differences can be more easily observed in red rectangles, which can be best viewed by zooming in.}}
\label{fig:impact_parameters}
\end{figure*}
%------------------------------------------------------------------------

%------------------------------------------------------------------------
\begin{figure}[t]
\centering
\setlength\tabcolsep{1.8pt}
\footnotesize
    \begin{tabularx}{\linewidth}{ccc}
        \subfigure{\includegraphics[width=0.32\linewidth]{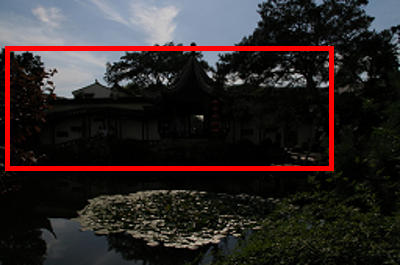}}&
        \subfigure{\includegraphics[width=0.32\linewidth]{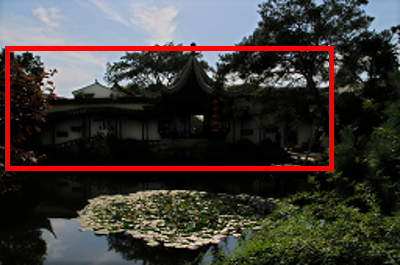}}&
        \subfigure{\includegraphics[width=0.32\linewidth]{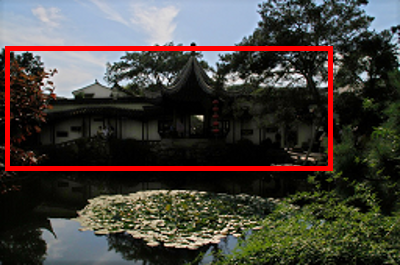}}
    \end{tabularx}
    \begin{tabularx}{\linewidth}{ccc}
        \subfigure{\includegraphics[width=0.32\linewidth]{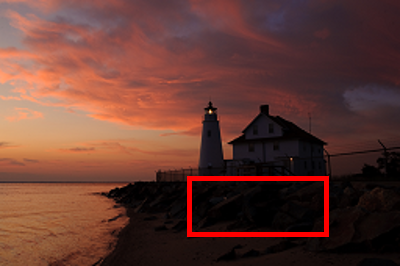}}&
        \subfigure{\includegraphics[width=0.32\linewidth]{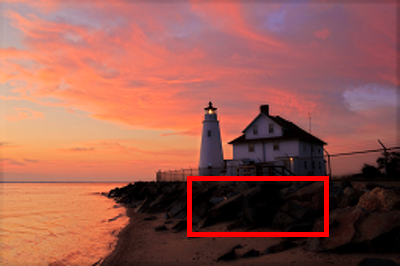}}&
        \subfigure{\includegraphics[width=0.32\linewidth]{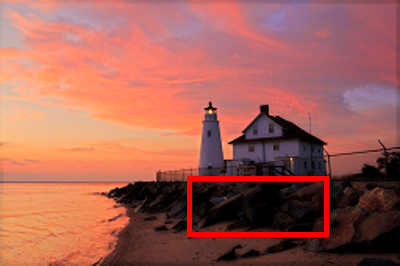}}
    \end{tabularx}
    \begin{tabularx}{\linewidth}{ccc}
        \subfigure{\includegraphics[width=0.32\linewidth]{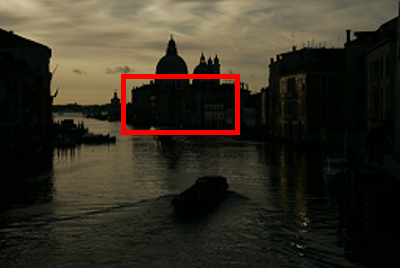}}&
        \subfigure{\includegraphics[width=0.32\linewidth]{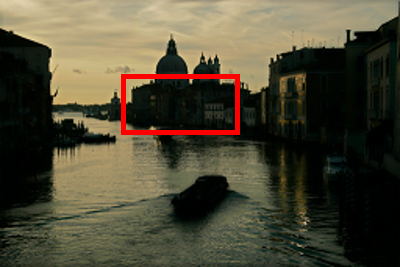}}&
        \subfigure{\includegraphics[width=0.32\linewidth]{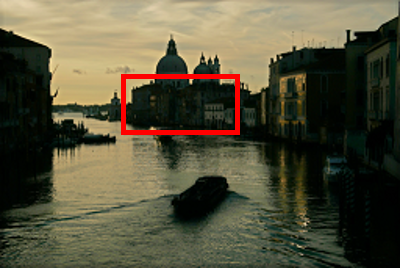}}\\
        Input image & w/o skip connection & IceNet\\
    \end{tabularx}
\caption{\modify{Impacts of the skip connection in the feature extractor. The images are enhanced at $\eta=0.6$ without scribbles.}}
\label{fig:impact_skip}
\end{figure}
%------------------------------------------------------------------------

For more subjective assessment, we asked the 11 participants to select the preferred image between the initially enhanced image and automatically enhanced ones obtained by SRIE \cite{fu2016weighted} and EnlightenGAN \cite{jiang2021enlightengan}. Note that SRIE and EnlightenGAN were the two most preferred algorithms among the conventional methods in the previous user study. Table~\ref{US:personal}(b) summarizes the voting results. The proposed algorithm won the most votes. Figure \ref{fig:qualitative_auto} compares qualitative results, in which the results of IceNet were obtained using the personalized exposure levels of a participant. The proposed IceNet provides more pleasing results without per-image interaction.

\subsubsection{Quantitative comparison}
To assess enhanced images quantitatively, we use 767 pairs of under- and normal-exposed images in the Part2 subset of the SICE dataset \cite{cai2018learning}, which was also adopted as the test set in \cite{guo2020zero}. Table \ref{table:QE} compares the average PSNRs and SSIMs on the Part2 subset. The scores of the conventional algorithms are excerpted from \cite{guo2020zero}, excluding LDR \cite{lee2013contrast}. \modify{We measured the average running times using 100 images of size $1200 \times 900$. For the conventional algorithms in \cite{lee2013contrast,guo2017lime,li2018structure}, we measured with a CPU because only the codes in the CPU versions were available.} For $\text{IceNet}^*$, we obtained the best score by controlling the exposure level $\eta$ for each image but without providing scribbles. For IceNet, we use initially enhanced images, described in Section~\ref{ssec:ice}. To determine the coefficients in Eq.~\eqref{eq:init} using the method of least squares, 70 pairs of $(y_i, \eta_i)$ were sampled during the test of $\text{IceNet}^*$. In Table \ref{table:QE}, we see that the proposed algorithm outperforms the conventional algorithms with large margins. Furthermore, IceNet provides outstanding performances even without per-image interaction.

\subsection{Ablation and Analysis}
\subsubsection{Preferences for contrast} In Figure~\ref{fig:userstudy}, the odd rows show how three participants enhanced the same input images differently during the user study and the even rows show that participants provided a variety of annotations to obtain satisfactory images. We see that the people have diverse preferences for contrast through the enhanced images. The proposed interactive algorithm can be used to satisfy these preferences adaptively.

\subsubsection{\modify{Impacts of parameter settings}} \modify{Figure \ref{fig:impact_parameters} compares the results of the proposed IceNet with different parameter settings. In this test, we change the number of convolutional layers and output channels. Note that the default setting of the proposed IceNet employs 7 convolutional layers with 32 output channels. As shown in Figure \ref{fig:impact_parameters}, when the amount of parameters is smaller than the default setting, the output images are over-enhanced. On the other hand, when the amount of parameters is larger than the default setting, the output images are similar to those of the default setting. This indicates that the default setting is both effective and efficient.
}

\subsubsection{\modify{Skip connection}} \modify{Figure \ref{fig:impact_skip} compares the results of IceNet \textit{with} and \textit{without} the skip connection. In this test, the output images are generated at $\eta=0.6$ without scribbles. It is observed that IceNet without the skip connection fails to bring out hidden details.}

\subsubsection{\modify{Loss function}}\label{sec:ablation_loss} \modify{Finally, we analyze the efficacy of each loss term in Eq.~\eqref{eq:overall_loss}. Figures \ref{fig:entropy}(b)$\sim$(e) show the results of IceNet trained with various combinations of the three losses. In this test, IceNet generates output images at $\eta=0.65$ without scribbles. First, we train IceNet without the interactive brightness control loss $\mathcal{L}_{\textrm{ibc}}$ in Figure \ref{fig:entropy}(c). By comparing it with Figure \ref{fig:entropy}(b), we see that removing $\mathcal{L}_{\textrm{ibc}}$ fails to enhance under-exposed regions. Second, we train IceNet without the entropy loss $\mathcal{L}_{\textrm{ent}}$ in Figure \ref{fig:entropy}(d). Compared to Figure \ref{fig:entropy}(b), the histogram is more concentrated, which indicates that the contrast is not improved sufficiently. Finally, Figure \ref{fig:entropy}(e) shows the result of IceNet trained without the smoothness loss $\mathcal{L}_{\textrm{smo}}$. Removing $\mathcal{L}_{\textrm{smo}}$ results in an over-enhanced image.}

\section{Conclusions}

We proposed IceNet for interactive CE, which is composed of several convolutional and fully-connected layers. IceNet enables a user to adjust image contrast easily. Specifically, a user provides an exposure level for controlling global brightness and a scribble map to darken or brighten local regions. Then, IceNet generates an enhanced image. The user may provide annotations iteratively until a satisfactory image is obtained. Also, IceNet produced a personalized enhanced image automatically, which can serve as a basis for further adjustment. Moreover, to train IceNet effectively and reliably, we developed the three differentiable loss functions. Extensive experiments on various datasets demonstrated outstanding CE performance of IceNet. \modify{Although IceNet is capable of providing a user with satisfactory results, the user should provide detailed scribbles. In the future, we will design a more user-friendly interactive CE method, which enable the user to control local brightness of desired regions more easily with simpler scribbles.}

\bibliographystyle{IEEEtran}
\bibliography{2021_Access_KSKO}

\clearpage

\begin{IEEEbiography}[{\includegraphics[width=1in,height=1.25in,clip,keepaspectratio]{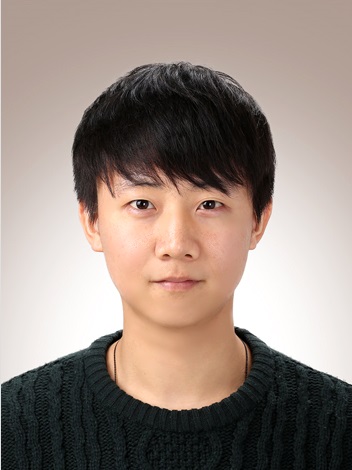}}]{Keunsoo Ko} (S'17) received the B.S. degree in electrical engineering from Korea University, Seoul, South Korea, in 2017, where he is currently pursuing the Ph.D. degree in electrical engineering. His research interests include image processing and machine learning.
\end{IEEEbiography}

\begin{IEEEbiography}[{\includegraphics[width=1in,height=1.25in,clip,keepaspectratio]{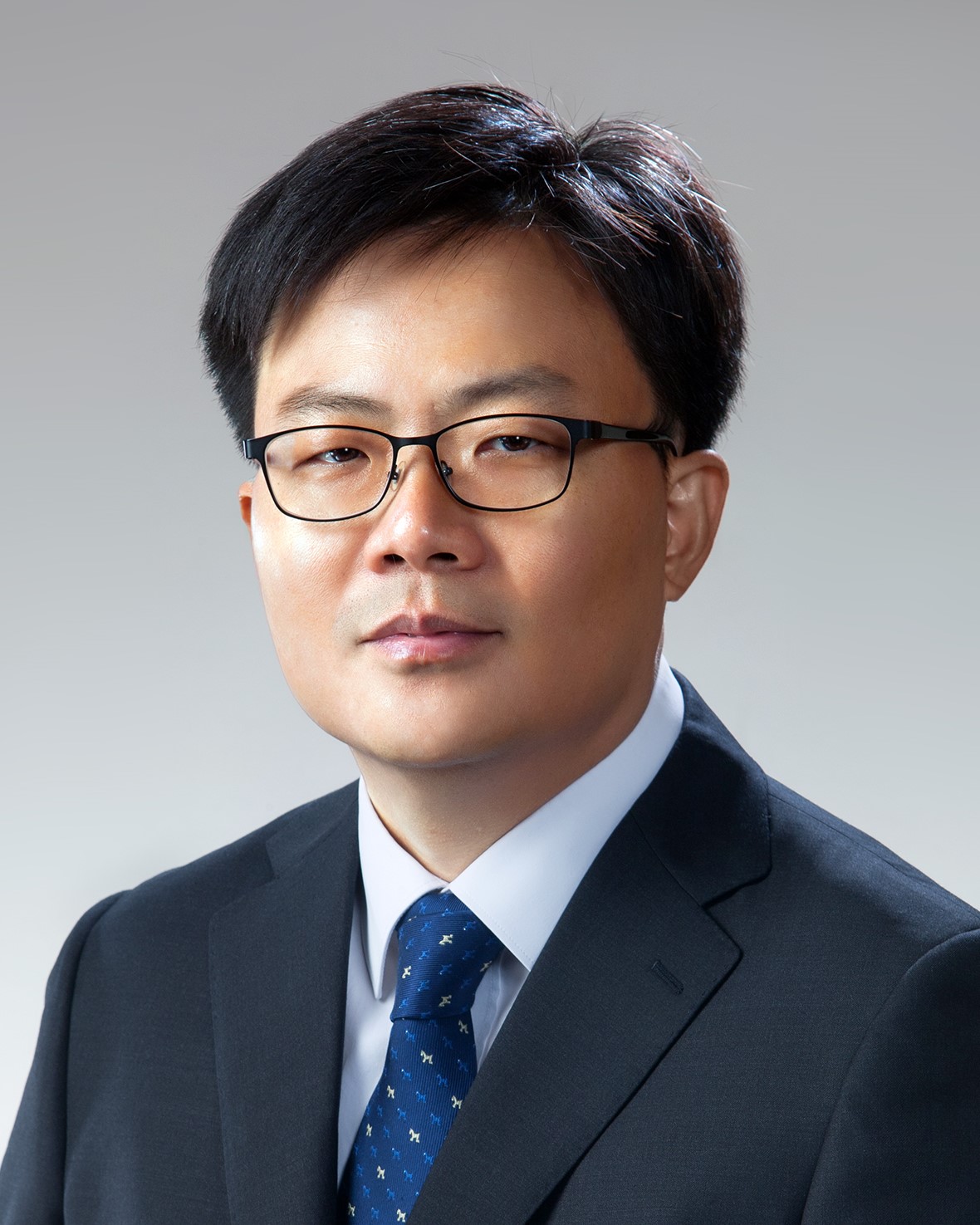}}]{Chang-Su Kim} (S'95-M'01-SM'05) received the Ph.D. degree in electrical engineering from Seoul National University with a Distinguished Dissertation Award in 2000. From 2000 to 2001, he was a Visiting Scholar with the Signal and Image Processing Institute, University of Southern California, Los Angeles. From 2001 to 2003, he coordinated the 3D Data Compression Group in National Research Laboratory for 3D Visual Information Processing in SNU. From 2003 and 2005, he was an Assistant Professor in the Department of Information Engineering, Chinese University of Hong Kong. In Sept. 2005, he joined the School of Electrical Engineering, Korea University, where he is a Professor. His research topics include image processing, computer vision, and machine learning. He has published more than 290 technical papers in international journals and conferences. In 2009, he received the IEEK/IEEE Joint Award for Young IT Engineer of the Year. In 2014, he received the Best Paper Award from Journal of Visual Communication and Image Representation (JVCI). He is a member of the Multimedia Systems \& Application Technical Committee (MSATC) of the IEEE Circuits and Systems Society. Also, he is an APSIPA Distinguished Lecturer for term 2017-2018. He served as an Editorial Board Member of JVCI and an Associate Editor of IEEE Transactions on Image Processing. He is a Senior Area Editor of JVCI and an Associate Editor of IEEE Transactions on Multimedia.
\end{IEEEbiography}

\EOD

\end{document}